\documentclass[twocolumn,trackchanges,twocolappendix]{aastex631}
\usepackage{ulem}
\usepackage{tabularx} 
\usepackage{booktabs}

\newcommand{\feka}{$\mathrm{Fe\:K\alpha}$}
\newcommand{\msun}{$M_{\odot}$}
\newcommand{\mch}{$\mathrm{M_{\mathrm{Ch}}}$}
\newcommand{\mdot}{$\dot{M}$}
\newcommand{\vwind}{$v_{\mathrm{wind}}$}
\newcommand{\twind}{$t_{\mathrm{wind}}$}

\begin{document}

\title{A Classification Scheme for X-ray Bright Type Ia Supernova Remnants Based on Their Circumstellar Interaction}

\author[0000-0003-3837-7201]{Travis Court}
\affiliation{Department of Physics and Astronomy, University of Pittsburgh, 3941 O'Hara Street, Pittsburgh, PA 15260, USA}
\affiliation{Pittsburgh Particle Physics, Astrophysics, and Cosmology Center (PITT PACC), University of Pittsburgh, Pittsburgh, PA 15260, USA}
\affiliation{Department of Physics and Astronomy, University of North Carolina Asheville, Asheville, NC 28804, USA}

\author[0000-0003-3494-343X]{Carles Badenes}
\affiliation{Department of Physics and Astronomy, University of Pittsburgh, 3941 O'Hara Street, Pittsburgh, PA 15260, USA}
\affiliation{Pittsburgh Particle Physics, Astrophysics, and Cosmology Center (PITT PACC), University of Pittsburgh, Pittsburgh, PA 15260, USA}

\author[0000-0002-2899-4241]{Shiu-Hang Lee}
\affiliation{Department of Astronomy, Kyoto University Oiwake-cho, Kitashirakawa, Sakyo-ku, Kyoto 606-8502 Japan}

\author[0000-0002-7507-8115]{Daniel Patnaude}
\affiliation{Center for Astrophysics | Harvard $\&$ Smithsonian, 60 Garden Street, Cambridge, MA 02138, USA}

\author[0000-0003-0894-6450]{Eduardo Bravo}
\affiliation{Departamento de F\'isica Te\'orica y del Cosmos, Universidad de Granada, E-18071 Granada, Spain}



\begin{abstract}
The parameter space for mass loss in Type Ia supernova progenitors is large, with different progenitor scenarios favoring different mass loss regimes. Here we focus on the impact that uniform and isotropic outflows have on the circumstellar environment of Type Ia supernova progenitors. We vary mass loss rate, wind velocity, and outflow duration, and evolve supernova remnant (SNR) models in this grid of circumstellar structures in order to compare the bulk properties of these models (ages, radii, and \feka\ centroids and luminosities) to observations. We find that roughly 55\% (7/13) of young X-ray bright Type Ia SNRs in the Milky Way and the Large Magellanic Cloud had progenitors that did not substantially modify their surroundings on $\sim$pc scales. This group includes SN Ia with a range of luminosities, and at least one likely product of a double detonation explosion in a sub-Chandrasekhar white dwarf. The other half of our sample can be divided in two distinct classes. A small subset of SNRs ($\sim$15\%, 2/13) have large radii and low \feka\ centroids and are likely expanding into large cavities that might have been excavated by fast ($\sim$1000 km/s), sustained progenitor outflows. The majority of the SNRs that are expanding into a modified medium ($\sim30\%$, 4/13) show evidence for dense material, likely associated with slow ($\sim$10 km/s) progenitor outflows, possibly a byproduct of accretion processes in near-Chandrasekhar white dwarfs spawned by younger stellar populations.




\end{abstract}



\section{Introduction} \label{sec:intro}
Type Ia supernovae (SNe) are foundational in our understanding of cosmology, but some of their fundamental properties remain obscure \citep[see][for reviews]{ruiter_type_2024,liu_type_2023,maoz_observational_2014}. Among these are the chain of events that leads to the SN Ia explosion itself, whereby a carbon-oxygen white dwarf (WD) in a binary system undergoes a thermonuclear runaway. One possibility is slow accretion of material from the companion, until either the WD mass gets close enough to the Chandrasekhar (\mch) limit that it becomes unstable and explodes on its own, or surface ignition of He-rich accreted material sends a shock wave towards the central regions of the WD that can trigger an ignition below the Chandrasekhar limit (this is often referred to as the double detonation scenario). Another possibility is a collision or merger between the WD and its companion either through dynamical secular evolution of triple systems \citep{Toonen2018}, by a rapid inspiral following a common envelope episode \citep{kashi_circumbinary_2011}, or through gravitational wave emission over long timescales in detached systems \citep{Webbink1984}. In a nutshell, carbon-oxygen WDs in SN Ia progenitors can have either non-degenerate or degenerate companions, they can explode close to or somewhat below \mch, and the accretion phase before the explosion can be long, or short, or non-existent. 


One thing that all evolutionary pathways for Type Ia SNe have in common is that they involve at least one phase of unstable mass transfer (i.e., a common envelope episode), as the orbital separations required for accretion or merger are orders of magnitude smaller than those of main sequence binaries (\citealt{wang_progenitors_2012,ivanova_common_2013}, but see also \citealt{ilkiewicz_wind_2019}). However, it is unclear how long such episodes last, when they take place, or how long after the episode the supernova explosion occurs \citep{ruiter_rates_2009,meng_common-envelope_2017,court_type_2024}. Mass transfer in binary systems is poorly understood, but it is highly unlikely that it would be conservative, and mass loss from the progenitor, whatever the mechanism behind it, has the potential to leave an imprint on the structure of its circumstellar material (CSM). 


Constraints on the density of the CSM around SN Ia progenitors have been derived from radio and X-ray follow-up campaigns of nearby SNe, weeks or months after the explosion. \cite{chomiuk_evla_2012} derived an upper limit of $\dot{M}/v_{\mathrm{wind}}\lesssim6\times10^{-10}\frac{M_{\odot}/yr}{100 km/s}$ or $n<6$ cm$^{-3}(\rho_{\mathrm{AM}}\approx1\times10^{-23}$ g/cm$^3)$ for SN 2011fe at radii between $\sim10^{15}-10^{16}$ cm. Similarly, 
\cite{margutti_no_2014} found similarly low mass loss rates $\dot{M}/v_{\mathrm{wind}}< 10^{-9}\frac{M_{\odot}/yr}{100 km/s}$ or $n<3$ cm$^{-3} (\rho_{\mathrm{AM}}\approx5\times10^{-24}$ g/cm$^3$) for SN 2014J at $R\sim 10^{16}$ cm. 
More broadly, \cite{chomiuk_deep_2016} found that $>94\%$ of SNe Ia should have mass loss rates below $\dot{M}/v_{\mathrm{wind}}\approx4\times10^{-7}\frac{M_{\odot}/yr}{100 km/s}$ at radii $\lesssim9\times10^{15}$ cm. Assuming a uniform ambient medium (AM) density, they found that
$>64\%$ of SNe Ia must be interacting with material of $n<100$ cm$^{-3}(\rho_{\mathrm{AM}}\approx1.67\times10^{-22}$ g/cm$^3$) at radii $\lesssim9\times10^{15}$ cm. 
Studies of the spectral evolution of large numbers of SNe have suggested that a small percentage of SN Ia might be interacting with much denser material. \cite{dubay_late-onset_2022} found that fewer than 5.1\% of Type Ias showed strong signs of interaction within 500 days of the explosion at radii $\lesssim9\times10^{16}$ cm, and fewer than 2.7\% between 500 and 1000 days at radii $\lesssim2\times10^{17}$ cm. These rare objects are commonly referred to as Ia-CSM SNe. Recently, \cite{Terwel2025} conducted a search for re-brightening events in $\sim7,000$ SN Ia on timescales of hundreds to thousands of days after the explosion, and found that such events are extremely rare, with only one confirmed event for the Ia-CSM SN 2020qxz.

X-ray observations of supernova remnants hundreds or thousands of years post-explosion allow us to probe the interaction between SN ejecta and CSM on larger spatial scales ($\sim10^{19}$ cm, or several pc), which are more relevant to the stellar evolution of SN Ia progenitors \citep[see][for a discussion]{patnaude_supernova_2017}. Bulk supernova remnant (SNR) properties like the centroid and luminosity of the \feka\ line, and the SNR radius and age, can be used to gauge the ability of specific models to reproduce the observations, and single out the most promising areas of the parameter space for SN Ia progenitor outflows. Previous work has shown that a non-modified, uniform AM with a range of densities similar to those found in the warm phase of the interstellar medium (ISM, $\rho_{\mathrm{ISM}}=0.04-5\times10^{-24} \mathrm{g/cm^3}$, corresponding to  $n=0.02-3.0$\ $\mathrm{cm^{-3}}$ -- \citealt{ferriere_interstellar_2001}) can provide a good approximation to the bulk properties of many Type Ia SNRs \citep{badenes_constraints_2006,badenes_are_2007,badenes_persistence_2008,yamaguchi_discriminating_2014,martinez-rodriguez_chandrasekhar_2018}. This does not rule out the presence of CSM in all cases - indeed, some SNRs do show signs of interaction with a modified CSM, like Kepler \citep{reynolds_deep_2007, chiotellis_imprint_2012, patnaude_origin_2012,katsuda_keplers_2015} and RCW 86 \citep{vink_x-ray_2006,badenes_are_2007,williams_rcw_2011,broersen_many_2014}, but it does showcase the importance of evaluating CSM models for SNRs quantitatively and in a consistent manner.


Here we present the first systematic investigation of the parameter space for CSM interaction in Type Ia SNRs, with fully coupled hydrodynamics and X-ray spectral calculations. We parametrize the progenitor mass loss history by varying the outflow velocity, \vwind, the mass loss rate, \mdot, and the outflow duration, \twind, producing CSM structures  with a large dynamic range in radius, density, and amount of mass injected into the ISM. We evolve SNR models in these structures and find that they also have a wide range of bulk properties, which only partially overlap with observations. This paper is organized as follows. The CSM outflow models are described in Section \ref{subsec:CSM}. The SN Ia explosion models and SNR models are described in Section \ref{subsec:Explosions}. In Section \ref{sec:results}, we discuss our results and compare the bulk properties of our models to observed SNRs. Lastly, in Sections \ref{sec:discussion} and \ref{sec:conclusion} we summarize our findings, outline our conclusions, and suggest avenues for future study.



\section{Methods} \label{sec:methods}
\subsection{Isotropic Outflow Models} \label{subsec:CSM}
\begin{figure*}
    \centering
    \includegraphics[width=\textwidth]{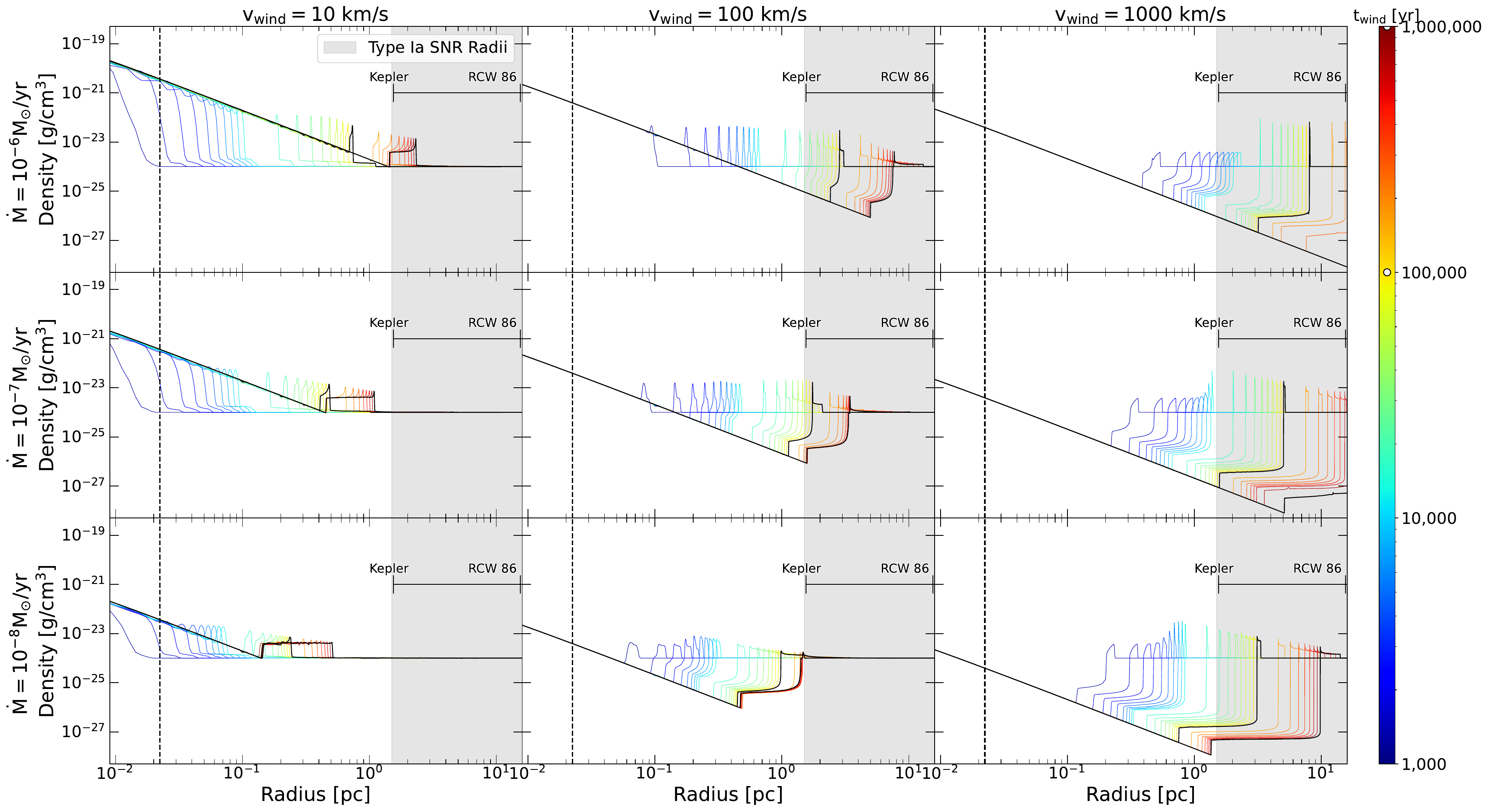}
    \caption{CSM structures sculpted by constant isotropic outflows, shown as density profiles as a function of radius, color coded by \twind. Each simulation spans $10^6$ years. The CSM is simulated to the size of the largest Type Ia SNRs, $\sim 15$ parsecs - the dynamic range of observed SNR sizes is indicated by the gray shaded region. The vertical dashed line corresponds to the outer layer of our \texttt{ddt24} explosion model after $10^7$ s of homologous expansion (see text for details). The top, middle, and bottom rows correspond to a mass loss rate of \mdot\ = $10^{-6},\ 10^{-7},\ 10^{-8}$ \msun/yr, respectively. The left, middle, and right columns correspond to \vwind\ = 10, 100, 1000 km/s, respectively. The density profiles used for the remnant simulations are overlayed in black at \twind\ values of 100,000 years and 1,000,000 years.}
    \label{fig:9_panel}
\end{figure*}


We use the 1D hydrodynamics code \texttt{VH-1} \citep{blondin_pulsar_2001,blondin_rayleigh-taylor_2001} to simulate the structure of the CSM around SN Ia progenitors. For the conditions of the ISM, we adopt $\rho_{\mathrm{ISM}}=10^{-24}$ g/cm$^3$ and $T_{\mathrm{ISM}}=10^4$ K, which result in $P_{\mathrm{ISM}}=8.3\times10^{-13}$ dyne/cm$^3$, assuming an ideal gas consistent with the warm phase of the ISM \citep{ferriere_interstellar_2001}. We assume that the progenitor ejects a continuous, uniform, and isotropic outflow that interacts with this ISM, leading to a density profile of the form $\rho_{\mathrm{AM}}(r)=Ar^{-2}$ close to the progenitor, where $A=\dot{M}/(4\pi v_{\mathrm{wind}})$ is the dilution parameter, \mdot\ is the mass loss rate in \msun/yr and \vwind\ is the speed of the outflow in km/s. Further away from the progenitor, there will be an interaction region between a reverse shock, which bounds the freely expanding outflow, and a forward shock propagating into the undisturbed ISM. The location, size, and internal structure of this interaction region will depend on the details of the outflow and the external pressure exerted by the ISM. \citep[see][for details]{castor_interstellar_1975,weaver_interstellar_1977,koo_dynamics_1992,koo_dynamics_1992-1}. Radiative cooling is taken into account using the cooling curves from \citealt{gnat_time-dependent_2007} and the script from \citealt{townsend_exact_2009}, assuming a solar metallicity ($Z=0.014$).



We produce the grid of CSM structures shown in Figure~\ref{fig:9_panel} by systematic variation of three parameters; \mdot, \vwind, and \twind. We choose three values of \mdot: $10^{-6},10^{-7},10^{-8}$ \msun/yr (top, middle and bottom rows), and three values of \vwind: 10, 100, 1000 km/s (left, middle, and right columns). Each outflow is simulated to $10^6$ years, with two snapshots considered, one at \twind\ $=10^5$ and another at $10^6$ years (color sequence).


Without making any specific assumptions about the pre-SN evolution of the progenitors, the variables that define our parameter space are meant to be representative of a wide range of physical scenarios. In a symbiotic binary, mass loss from the nondegenerate star would be slow, \vwind $\lesssim$ 100 km/s, with mass loss rates between $2\times10^{-9}-2\times10^{-6}$\mdot/yr \citep{chen_tidally_2011}. Mass lost at the outer Lagrangian point should escape with velocities of a few 100 km/s (see \citealt{margutti_no_2014} for references). Lastly, mass loss through optically thick accretion winds would be close to the escape velocity of the white dwarf $\sim1000$ km/s \citep{hachisu_new_1996,Prinja2000,Cuneo2023}. While these values encompass much of the relevant parameter space for the CSM around SN Ia progenitors, some outflow regimes are outside the scope of the present work. Specifically, we do not consider episodic or anisotropic outflows, which are expected theoretically \citep{theuns_wind_1993,wood-vasey_novae_2006} and present in real binary systems involving a white dwarf (like U Sco and RS Oph, see \citealt{Diaz2010,booth_modelling_2016}), although unambiguous observational evidence connecting these outflows with SN Ia progenitors on large scales has remained elusive \citep{cendes_thirty_2020}.

\begin{figure}
    \centering
    \includegraphics[trim= 10 20
    0 10, clip,
    width=\columnwidth]{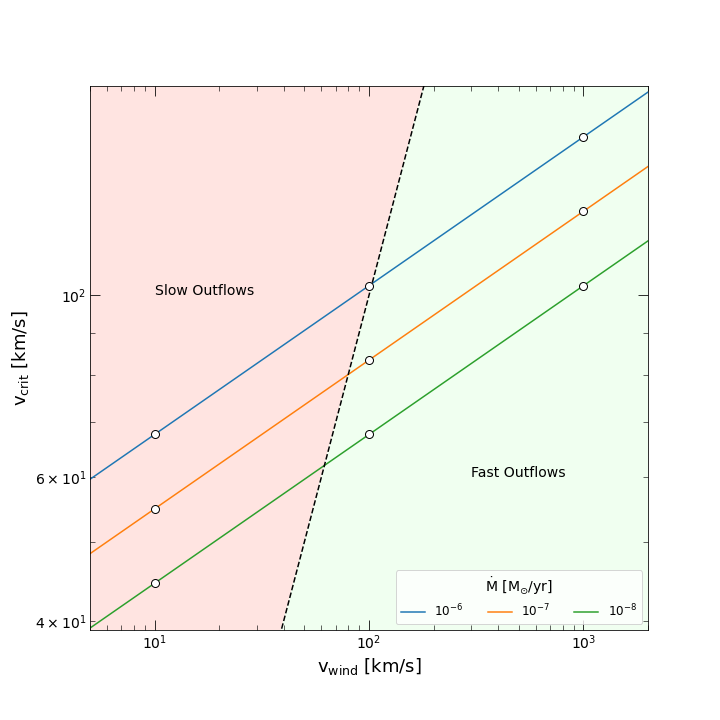}
    \caption{Critical velocity for our isotropic outflow models (white circles). The solid colored lines indicate different mass loss rates, while the dashed black line shows the $v_{\mathrm{crit}}=v_{\mathrm{wind}}$ boundary that divides, slow, momentum driven outflows from fast, energy-driven outflows \citep{koo_dynamics_1992}.}    
    \label{fig:critical velocity}
\end{figure}

These CSM structures can be divided into two broad groups, depending on whether the shocked outflow undergoes radiative losses or not: momentum driven structures and energy driven structures. The dividing line is the critical velocity, $v_{\mathrm{crit}}$, given by:

\begin{equation}
    v_{\mathrm{crit}}=10^4 \left(\frac{\dot{M}\,v_{\mathrm{wind}}^2}{2}\frac{\rho_{\mathrm{ISM}}}{\mu_H}\right)^{1/11}\ \mathrm{cm/s}
    \label{eq:v_crit}
\end{equation}
where $\mu_H=2.34\times 10^{-24}$ g is the mean mass per H atom in a gas with solar abundances \citep{koo_dynamics_1992}. In our model grid, CSM structures generated with \vwind\ $=10$ km/s are momentum driven, and those generated by faster outflows are energy driven (see Figure~\ref{fig:critical velocity}). Note that the model with \vwind\ $=100$ km/s and \mdot\ $=10^{-6}$ \msun/yr sits on the $v_{\mathrm{crit}}=v_{\mathrm{wind}}$ boundary, but it behaves like an energy driven outflow when compared to other outflows in Figure \ref{fig:9_panel}. At \twind\ $=10^6$ yr the momentum driven outflows (left column of Figure~\ref{fig:9_panel}) feature a more or less smooth transition between the $\rho \propto r^{-2}$ freely expanding wind and the ISM, at a radius that varies between a fraction of a pc for the lowest \mdot\ outflows and a few pc for the highest \mdot. By contrast, energy driven outflows lead to cavities, with densities orders of magnitude lower than the ISM and sizes of $\sim$1 pc for \vwind\ $=100$ km/s and $\sim$10 pc for \vwind\ $=1000$ km/s, with larger \mdot\ models producing larger cavities. In most cases, a dense shell of radiatively cooled material appears at the contact discontinuity between shocked outflow and shocked ISM. These CSM structures are similar in size and structure to the ones calculated by \citealt{badenes_are_2007}. The main difference is that \citealt{badenes_are_2007} tied their outflow properties to specific evolutionary models for SN Ia progenitors, including varying \mdot\ and in some cases mass-conservative phases before the SN explosion, while we restrict ourselves to uniform and continuous outflows.  

Other than varying the outflow parameters, there are two ways to alter these CSM structures. One is changing the external pressure exerted by the ISM. In general, larger ISM pressures (i.e., higher $\rho_{\mathrm{ISM}}$ or $T_{\mathrm{ISM}}$) will  result in smaller and denser CSM structures, and vice versa. 
The other is changing the metallicity, which will impact the radiative cooling. In test runs using a sub-solar metallicity ($Z=0.1Z_{\odot}$), momentum driven CSM structures have larger radii, up to 1 pc in the most extreme cases at $t_{\mathrm{wind}}=10^6$ yr. At $t_{\mathrm{wind}}=10^5$ yr, CSM structures become a fraction of a parsec smaller. 
The energy driven structures show no significant differences at lower metallicity. For the reminder of this work, we adopt solar metallicity for our CSM structures and maintain $\rho_{\mathrm{ISM}}=10^{-24}$ g/cm$^3$ and $T_{\mathrm{ISM}}=10^{4}$ K. 

We note that the CSM structures shown in Figure~\ref{fig:9_panel} span a larger dynamic range in radii than the known Type Ia SNRs listed in \citealt{martinez-rodriguez_chandrasekhar_2018}, which are between $\sim$2 pc (Kepler and G1.9+0.3 \citealt{reynolds_deep_2007,borkowski_nonuniform_2014}), and $\sim$16 pc (RCW 86,  \citealt{williams_rcw_2011,broersen_many_2014}). For most of the SNRs in this sample, the forward shock would have overrun most or all of the CSM produced by any slow outflows with parameters similar to the ones we consider here.

\subsection{Type Ia Explosion Models} \label{subsec:Explosions}

\citealt{martinez-rodriguez_chandrasekhar_2018} showed that the ambient medium interaction has a larger impact on the bulk properties of Type Ia SNRs than the details of the explosion model. Given the large parameter space for CSM interaction that we consider here, we have chosen to use a single explosion model for this study: a CO WD with a mass near \mch $= 1.4$ \msun\ that undergoes a thermonuclear runaway in its central regions and explodes, with a burning front that undergoes a deflagration to detonation at a specific density \citep{khokhlov_delayed_1991}. We use an intermediate energy explosion of this subclass, model \texttt{ddt24}  from \cite{bravo_snr-calibrated_2019}. The deflagration to detonation density in this model is $\rho_{\mathrm{DDT}}=2.4\times10^7$ g/cm$^3$, with a central density of $\rho_c=3.0\times10^9$ g/cm$^3$, an explosion energy of $E_K=1.43\times10^{51}$ erg, a total Fe yield of $0.80$\msun\ and a synthesized mass of $0.70$\msun\ of $^{56}$Ni, representative of a canonical SN Ia \citep{stritzinger_consistent_2006,scalzo_ejected_2014}.



\subsection{Supernova Remnant Models and Synthetic Spectra}

The CSM structures and SN Ia explosion model are input as initial conditions into \texttt{ChN}, and their interaction (i.e., the SNR model) is followed to an age of 5000 years. \texttt{ChN} is a multipurpose code that combines hydrodynamics (HD), non-equilibrium ionization (NEI), plasma emissivities, radiative cooling, and forbidden line emission \citep{ellison_particle_2007,lee_generalized_2012,lee_cr-hydro-nei_2013,lee_reverse_2014,lee_modeling_2015,patnaude_role_2009,patnaude_role_2010,court_type_2024}. 
Although \texttt{ChN} has the ability to account for the effect of cosmic ray acceleration on the SNR dynamics \citep{patnaude_role_2009,patnaude_role_2010}, we have chosen not to include this parameter in our simulations. While cosmic ray acceleration has a measurable and well-characterized impact on the dynamics of the forward shock and the thermal emission from the shocked AM \citep{decourchelle_thermal_2000,warren_cosmic-ray_2005}, the
effects on the reverse shock dynamics and thermal emission from the shocked ejecta are not as well understood \citep{badenes_constraints_2006,badenes_persistence_2008,yamaguchi_new_2014}. 
We will return to the potential impact of cosmic ray acceleration in Section~\ref{subsec:outside} below. 

We initialize each simulation by homologously expanding the SN ejecta for $10^7$ s and appending our calculated CSM structure to the outermost radius of ejecta, which at this time is located at $6.9\times10^{16}$ cm. The position of the outermost layer of ejecta after this expansion is indicated in Figure \ref{fig:9_panel} with a vertical dashed line. This phase of homologous expansion is necessary to ensure model convergence, and the mass of CSM removed by this procedure is negligible in all cases: $<8\times10^{-3}$ \msun, less than $1\%$ of the ejecta mass. For the \vwind\ $=1000$ km/s with \mdot\ $=10^{-6}$ and $10^{-7}$ \msun/yr, we had to increase the homologous expansion phase to $3.15\times10^{7}$ s (1 year), as the criteria for convergence in these low-denisty models are more stringent. This leads to an ejecta radius of $1.8\times10^{17}$ cm, while still removing an amount of CSM equivalent to less than $1\%$ of the ejecta mass. The longer homologous expansion phase used in these models does not significantly impact the thermal history of the SNR.


Besides the CSM structure and SN Ia explosion model, there is one additional parameter in our SNR simulations: the amount of collisionless electron heating at the reverse shock \citep{badenes_thermal_2005,yamaguchi_new_2014}. The value of this parameter is not known from first principles \citep[see][]{ghavamian_physical_2007}. In the absence of collisionless heating, the temperature of the species (ions or electrons) downstream from the shock is given by 
\begin{equation}
    T_{\mathrm{i,e}} = \frac{3}{16}\frac{m_{\mathrm{i,e}}v_s^2}{k_{\mathrm{b}}}
\end{equation}
where $m_{i,e}$ is the mass of the ions and electrons, $v_s$ is the shock speed, and $k_{\mathrm{b}}$ is the Boltzmann constant. Multiple lines of evidence \citep[see][and references therein]{yamaguchi_new_2014} indicate that the electrons are heated above this minimum temperature in the reverse shocks of young SNRs, which can have important consequences for the X-ray emission from the shocked ejecta \citep{badenes_thermal_2005}. We parameterize the efficiency of this heating as the ratio of the post shock temperatures
\begin{equation}
    \beta \equiv \frac{T_{\mathrm{e}}}{T_{\mathrm{i}}}
    \label{eq:beta_long}
\end{equation}
where $T_e$ and $T_i$ are the temperatures of the electron and ion populations, respectively. Within \texttt{ChN}, this is parametrized as multiples of the electron to proton mass ratio, $m_e/m_p=5.45\times10^{-4}$, so that heating can be applied across elemental populations. For reference, we consider the value for Fe, $\beta_{\mathrm{min}}=\frac{m_e}{55.8m_p}\approx1\times10^{-5}$. We further discuss this in Section \ref{sec:beta}.





For each SNR simulation, we calculate synthetic X-ray spectra using the \texttt{NEISession} package within the \texttt{pyatomdb} module \citep{foster_pyatomdb_2020}, as described in \citealt{court_type_2024}. \texttt{ChN} outputs all the necessary data for this step: temperatures for the electron and ion species, density, chemical composition, and ionization state for each Eulerian layer in the SNR model. These raw spectra can be convolved with any instrument response. When discussing our results, we do not include the thermal emission from the shocked AM, as it makes a small contribution to the integrated X-ray spectrum in the objects we discuss here, and a negligible contribution to the flux in the \feka\ line. See the Appendix of  \citealt{court_type_2024} for a description of our procedure to calculate \feka\ line centroids and fluxes from synthetic spectra.



\section{Results} \label{sec:results}


\subsection{SNR models in a uniform AM}
\label{sec:uniformAM}

\begin{table*}
    \centering
    \caption{Observational properties of Type Ia SNRs}
    \begin{tabular}{l c c c c c c c}
    \hline \hline
        SNR Name  & $\mathrm{E_{FeK\alpha}}$ & $\mathrm{F_{FeK\alpha}}$ & Distance & $\mathrm{L_{FeK\alpha}}$ & Radius & Age & References \\ 
        & [eV] & [$\mathrm{10^{-5}\ ph\ cm^{-2}\ s^{-1}}$] & [kpc] & [$\mathrm{10^{40}\ ph\ s^{-1}}$] & [pc] & [yr] & \\ 
        \hline
        G1.9+0.3& 6444 & 0.12 & $\sim8.5$ & 1 & $\sim2.0$ & $\lesssim150$ & (1), (2) \\
        0509-67.5& $6425^{+14}_{-15}$ & $0.32\pm0.04$ & 50 & $96\pm12$ & 3.6 & $\sim400$ & (3) \\
        Kepler & $6438\pm1$ & $34.6\pm0.2$ & $3.0-6.4$ & $91\pm66$ & $2.3\pm0.9$ & 421 & (4) \\
        Tycho & $6431\pm1$ & $61.0\pm0.4$ & $3.2^{+0.1}_{-0.2}$ & $75_{-9}^{+5}$ & $3.9_{-0.2}^{+0.1}$ & 453 & (5) \\
        0519-69.0& $6498^{+6}_{-8}$ & $0.93\pm0.05$ & 50 & $278\pm15$ & 4.0 & $\sim600$ & (3) \\
        N103B& $6545\pm6$ & $2.15\pm0.10$ & $50$ & $643\pm30$ & 3.6 & $\sim860$ & (3) \\
        SN 1006 & $6429\pm10$ & $2.55\pm0.43$ & $2.2$ & $1.5\pm0.3$ & 10 & 1019 & (6) \\
        G352.7-0.1& $6443^{+8}_{-12}$ & $0.82\pm0.08$ & $10.5$ & $10.8\pm0.5$ & $8.4$ & $\sim1600$ & (7), (8) \\
        RCW 86 & $6408^{+4}_{-5}$ & $14.0\pm0.7$ & $2.5$ & $10.5\pm0.5$ & $16$ & 1840 & (9) \\
        3C 397 & $6556^{+4}_{-3}$ & $13.7\pm0.4$ & $6.5-9.5$ & $105\pm39$ & $5.3\pm0.5$ & $1350-5300$ & (10) \\
        DEM L17& $6494\pm58$ & $0.09^{+0.02}_{-0.03}$ & 50 & $26^{+8}_{-9}$ & 8.6 & $\sim4700$ & (11) \\
        G344.7-0.1& $6463^{+9}_{-10}$ & $4.03\pm0.33$ & $6.3$ & $19\pm8$ & $7.2$ & $3000-6000$ & (12) \\
        G337.2-0.7 & $6505^{+26}_{-31}$ & $0.21\pm0.06$ & $2.0-9.3$ & $0.8\pm1.1$ & $4.9\pm3.2$ & $5000-7000$ & (13) \\
        \hline
    \end{tabular}
    \tablecomments{Observational properties of the Type Ia SNRs listed by \cite{martinez-rodriguez_chandrasekhar_2018}. \feka\ centroids and fluxes, and SNR angular radii are taken from \cite{yamaguchi_discriminating_2014}, except for G1.9+0.3 \citep{borkowski_supernova_2013} and DEM L71 \citep{maggi_population_2016}. The distances used to calculate the \feka\ luminosities and SNR radii for the Galactic SNRs, and the age estimates for the non-historic SNRs are taken from the listed references: (1) \cite{reynolds_youngest_2008}; (2) \cite{borkowski_supernova_2013}; (3) \cite{rest_light_2005}; (4) \cite{reynoso_new_1999}; (5) \cite{neumann_echo_2024}; (6) \cite{yamaguchi_x-ray_2008}; (7) \cite{zhang_molecular_2023}; (8) \cite{pannuti_xmm-newton_2014}; (9) \cite{helder_proper_2013}; (10) \cite{leahy_distance_2016}; (11) \cite{hughes_iron-rich_2003}; (12) \cite{fukushima_element_2020}; (13) \cite{rakowski_can_2006}.
    }
    \label{tab:TypeIaSNRs}
\end{table*}

Before we discuss SNR models with CSM interaction, we briefly comment on the salient features of Type Ia SNR models interacting with a uniform AM \citep[see][for more detailed discussions]{badenes_constraints_2006,badenes_persistence_2008, yamaguchi_discriminating_2014,martinez-rodriguez_chandrasekhar_2018}. We calculated SNR models interacting with seven values of a uniform AM density: 0.04, 0.1, 0.2, 0.3, 1.0, 2.0, and 5.0$\times10^{-24}$ g/cm$^3$, which span the typical range found in the warm phase of the ISM in the Milky Way \citep{ferriere_global_1998,berkhuijsen_density_2008}. 

\begin{figure}
    \centering
    \includegraphics[width=\columnwidth]{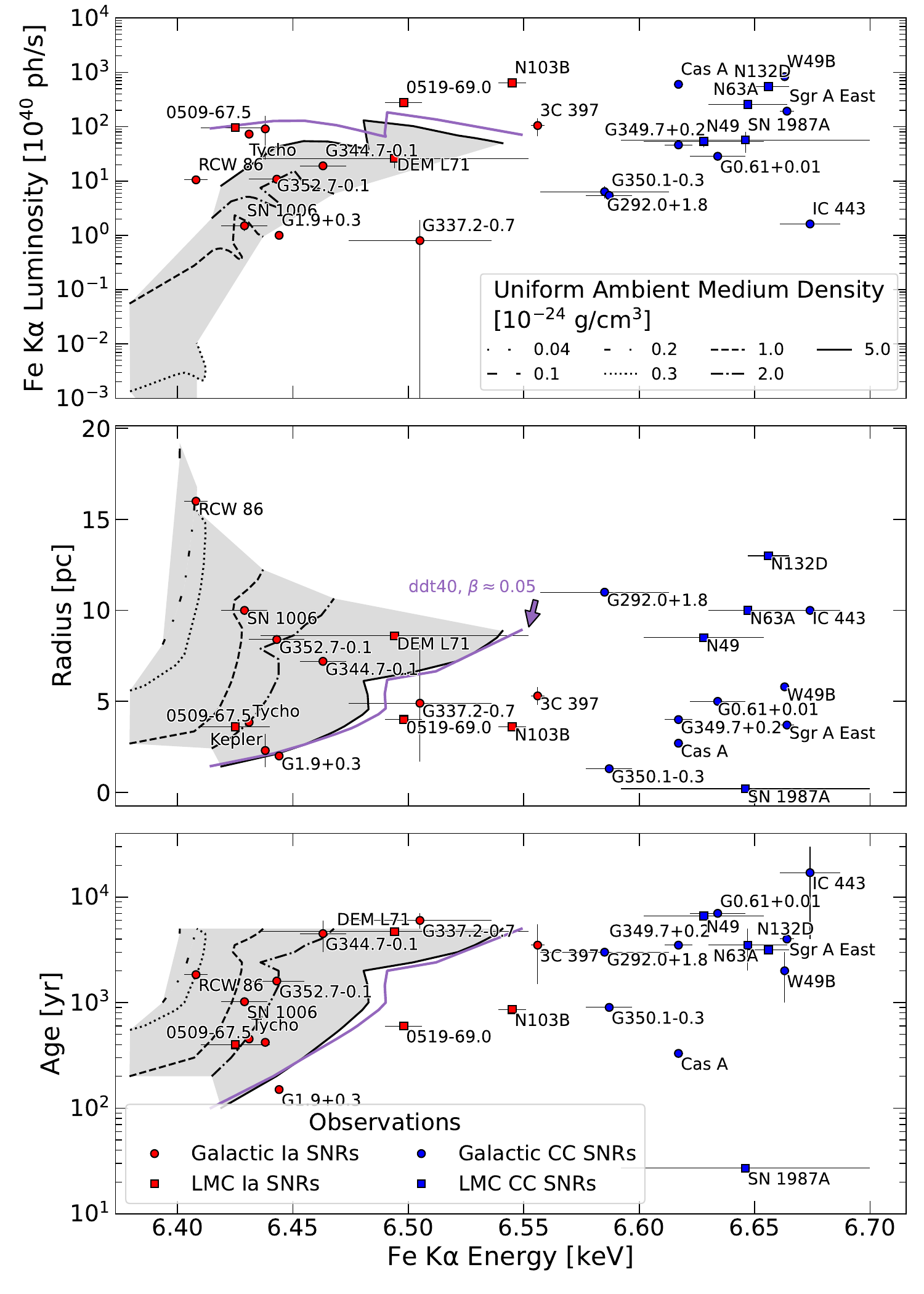}
    \caption{Comparison between SNR models interacting with a uniform AM and SNR observations. Bulk properties are shown as a function of \feka\ centroid energy: \feka\ luminosity (\textit{top}), forward shock radius (\textit{middle}), and remnant age (\textit{bottom}). The shaded region highlights the area of parameter space covered by the uniform ambient medium models with the line style representing the density. The purple overlaid line corresponds to a more energetic explosion model (\texttt{ddt40} - see text for details). Observed values for Type Ia SNRs are shown with red symbols, while Core Collapse SNRs are shown with blue symbols. The shape of these symbols (circles vs. squares) distinguishes Milky Way from LMC SNRs. } 
    \label{fig:uniform}
\end{figure}

In Figure~\ref{fig:uniform} we evaluate these uniform AM models using four key observable diagnostics of the bulk dynamics of SNRs: the centroid of the \feka\ line blend, which is a proxy for the ionization timescale in the shocked ejecta \citep{yamaguchi_discriminating_2014}, as well as the \feka\ luminosity, the SNR radius, and the SNR age. We compare the values predicted by the models with the observational properties of a sample of SNRs with \feka\ emission in the Milky Way and the Large Magellanic Cloud adapted from the compilations in \cite{yamaguchi_discriminating_2014} and \cite{martinez-rodriguez_chandrasekhar_2018}, with a few updates and modifications - see Table~\ref{tab:TypeIaSNRs} for details. At this stage, our comparisons are general and qualitative -- in Section~\ref{subsec:CompObs} we will discuss specific cases in further detail. It is by now well established that most Type Ia SNRs are clustered in these \feka\ luminosity, radius, and age vs. \feka\ centroid plots, and cleanly separated from most CC SNRs by the value of the \feka\ centroid. In all Type Ia SNRs with a secure classification, this value is below $\sim$6.55 keV, which corresponds to a charge state between 20 and 21, and an emission measure averaged ionization timescale of $\sim5\times10^{10}$ $\mathrm{cm^{-3}\,s}$ in the shocked ejecta -- see \citealt{yamaguchi_new_2014}. We emphasize that we are using these plots as a broad diagnostic tool, and not as a means to type individual SNRs -- see \cite{maggi_fe_2017}, \cite{siegel_can_2021}, \cite{dang_typing_2024} for a discussion and caveats regarding individual objects.

Our results are in qualitative agreement with previous studies, in that the bulk dynamics and spectral properties of uniform AM models are a good approximation to the measured properties of most Type Ia SNRs, particularly for the higher values of $\rho_{\mathrm{AM}}$. 
Quantitatively, the main difference between our results and those of \citealt{martinez-rodriguez_chandrasekhar_2018} is that our \feka\ luminosities are smaller by a factor of $\sim2-5$ for SNR models with similar parameters. This discrepancy might originate in our use of the Eulerian version of \texttt{ChN} instead of the Lagrangian version used by \cite{martinez-rodriguez_chandrasekhar_2018}. We require an Eulerian code because the parameter study we present here includes models with strong CSM interaction \citep[see][for a discussion]{court_type_2024}. It is possible that the loss of resolution in some of the densest layers of shocked ejecta incurred by our use of an Eulerian code results in a lower predicted \feka\ luminosity for these models. In any case, we note that our \feka\ centroids (and therefore, the emission measure averaged ionization timescales in the shocked ejecta) are consistent with the values obtained using the Lagrangian version of \texttt{ChN}. Other features of the 1D models, like the change in regime of the evolution of the \feka\ centroid caused by the reverse shock bounce, which is clearly seen at an age of $\sim10^3$ yr for the highest AM densities, are also in agreement with Lagrangian calculations.


We note that it is possible to extend the parameter space covered by uniform AM models towards larger \feka\ luminosities by using more energetic SN Ia models and introducing collisionless electron heating at the reverse shock (see Section~\ref{sec:beta}). Also shown in Figure~\ref{fig:uniform} is a SNR model with $\rho_{\mathrm{AM}}=5.0\times10^{-24}$ g/cm$^3$ and $\beta=0.05$, calculated using the SN Ia model \texttt{ddt40} from \citealt{bravo_snr-calibrated_2019}. This explosion model has a higher deflagration to detonation density ($\rho_{\mathrm{DDT}}=4.0\times10^7$ g/cm$^3$) than our fiducial model \texttt{ddt24}, which leads to a higher Fe yield (0.97 \msun vs 0.80 \msun), with the layers dominated by Fe extending further out in the SN ejecta. This change in Fe content and distribution, together with the increased collisionless electron heating at the reverse shock, results in a \feka\ luminosity an order of magnitude higher at early SNR ages. We will discuss the effect of collisionless electron heating in greater detail in Section~\ref{sec:beta} below.



\subsection{SNR models in slow progenitor outflows}
\label{sec:slow}

\begin{figure}
    \centering
    \includegraphics[width=\columnwidth]{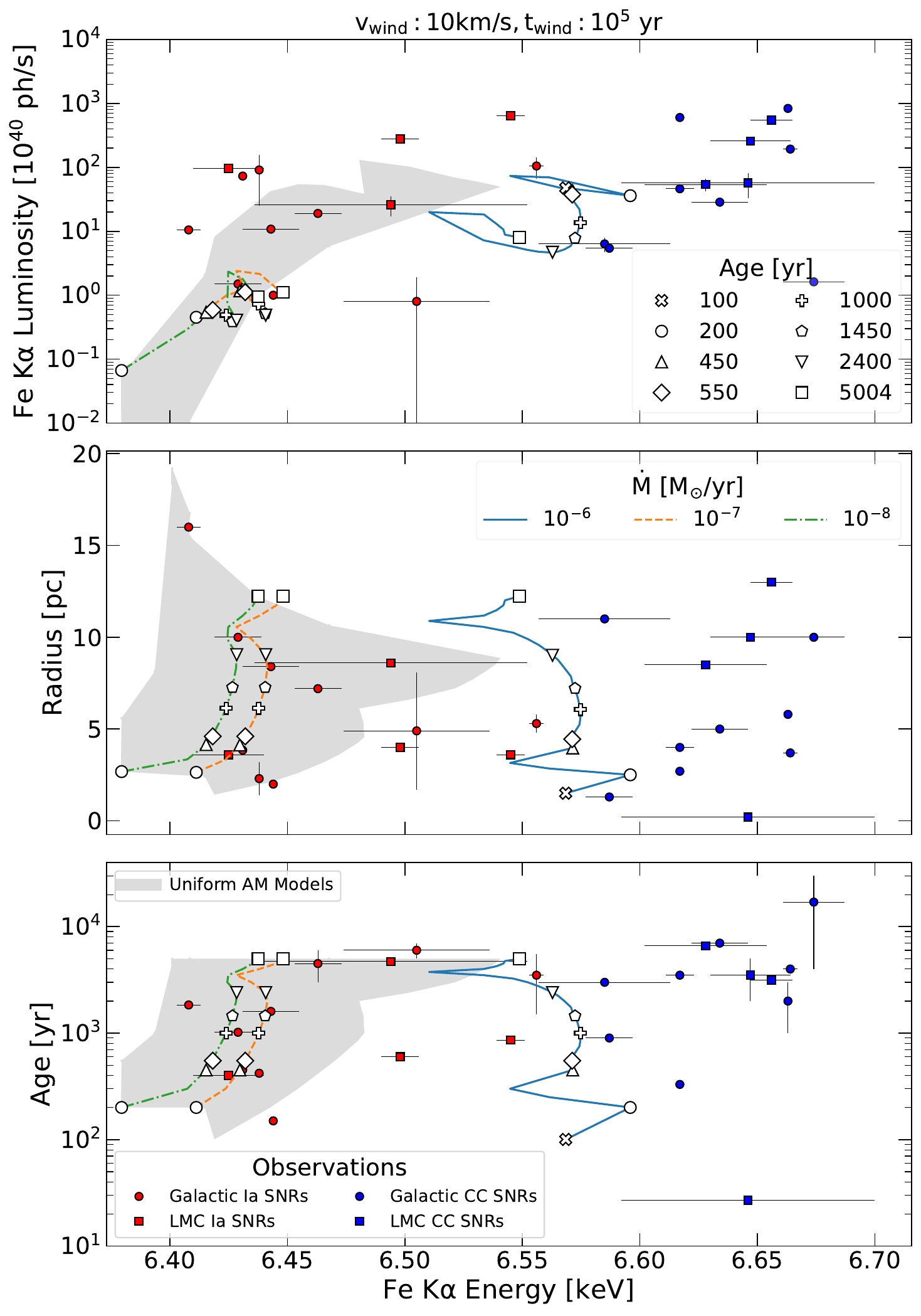}
    \caption{Comparison between SNR models interacting with slow outflows ($v_{\mathrm{wind}}=10$ km/s) and SNR observations.  Bulk properties are shown as a function of \feka\ centroid energy: \feka\ luminosity (\textit{top}), forward shock radius (\textit{middle}), and remnant age (\textit{bottom}). The dashed-dotted green line corresponds to $\mathrm{\dot{M}=10^{-8}\ M_{\odot}/yr}$, the dashed orange line corresponds to $\mathrm{\dot{M}=10^{-7}\ M_{\odot}/yr}$, and the solid blue line corresponds to $\mathrm{\dot{M}=10^{-6}\ M_{\odot}/yr}$, with empty symbols indicating specific SNR ages. Observed values for Type Ia SNRs are shown with filled red symbols, while Core Collapse SNRs are shown with filled blue symbols. The shape of these symbols (circles and squares) distinguishes Milky Way from LMC SNRs. A shaded region corresponding to the parameter space spanned by the uniform $\mathrm{\rho_{\mathrm{AM}}}$ models is included for comparison.}
    \label{fig:slow_1e5yr}
\end{figure}

The parameter space covered by SNR models interacting with slow outflows (\vwind\ $=10$ km/s, \twind\ $=10^5$ yr) is shown in Figure~\ref{fig:slow_1e5yr}. By the time there is enough shocked Fe emission in these SNR models to show up in our plots (i.e., the synthetic spectra have enough flux in the \feka\ line above the continuum to calculate a centroid), the forward shock has often overrun the CSM structure, and is already interacting with the ISM. The slow outflow models with the lowest \mdot\ ($10^{-8}$ and $10^{-7}$ \msun/yr) are therefore very similar to the uniform AM model with $\rho_{\mathrm{AM}} = 10^{-24}$ g/cm$^3$. In these models, we do not measure substantial ionization before a SNR age of 450 years, and even at the latter ages the \feka\ luminosities are two to three orders of magnitude below the brightest Type Ia SNRs. We note that these problems could be solved by increasing the value of the ISM density outside the CSM structure, but that would be no different than varying $\rho_{\mathrm{AM}}$ as done in Section~\ref{sec:uniformAM}, essentially decoupling the SNR models from the properties of the progenitor outflow.

As \mdot\ increases to $10^{-6}$ \msun/yr, the models become more luminous and more highly ionized at a given age, reflecting the fact that the SN ejecta has interacted with more dense material in the past (see Figure~\ref{fig:9_panel}), but the SNR radii do not change much. This can be partially explained by the fact that the interaction with the densest CSM material occurs within the first few hundred years of SNR evolution, and within a parsec of the progenitor, when the forward shock is fastest and more difficult to decelerate. At this \mdot, the \feka\ centroid energies are already higher than those observed in the most highly ionized Type Ia SNRs, and close to the region of parameter space inhabited by CC SNRs, with \feka\ luminosities comparable to the highest values produced by uniform AM models.


\subsection{SNR models in fast progenitor outflows}
\label{sec:fast}

\begin{figure*}
    \centering
    \includegraphics[width=\columnwidth]{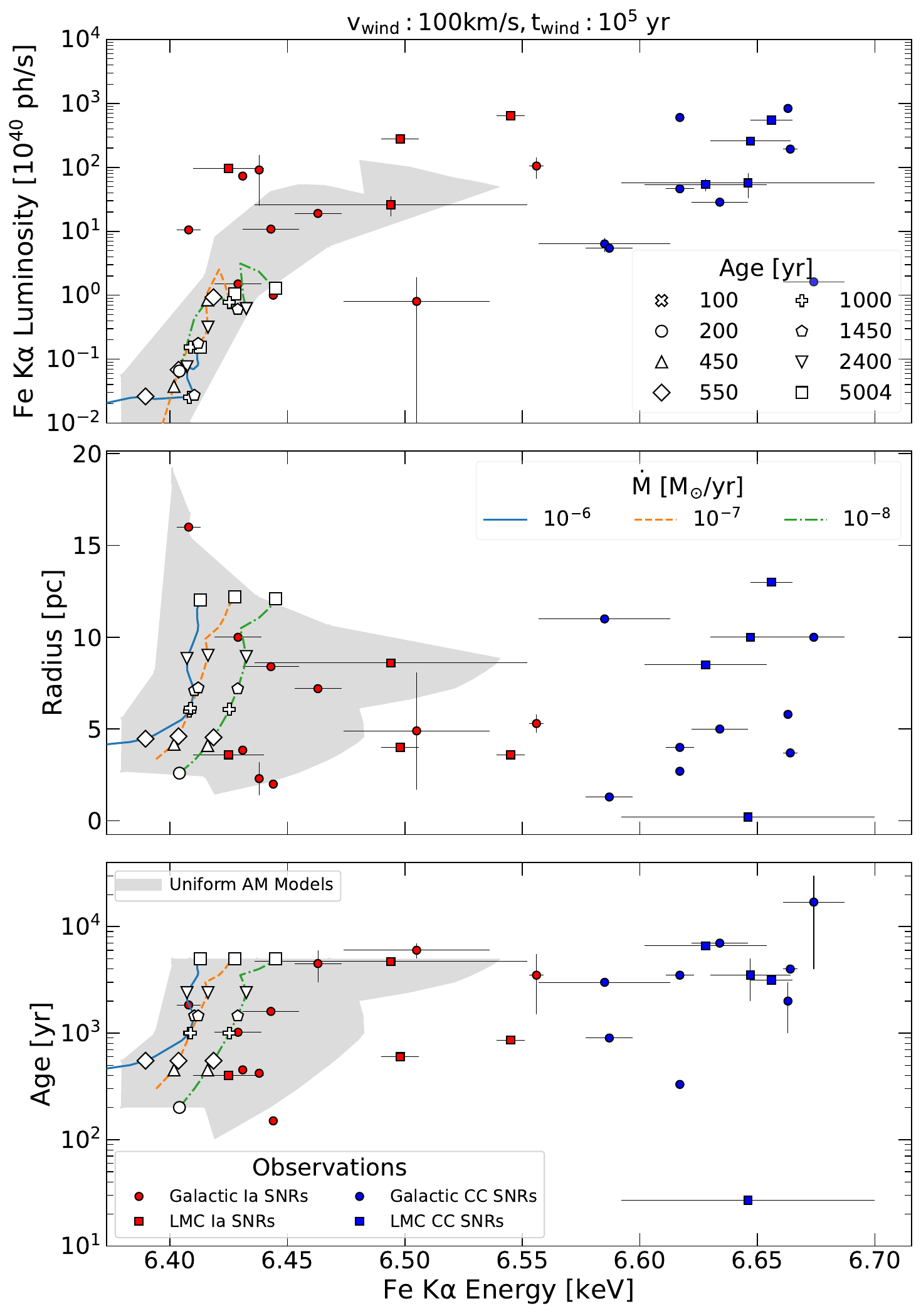}\includegraphics[width=\columnwidth]{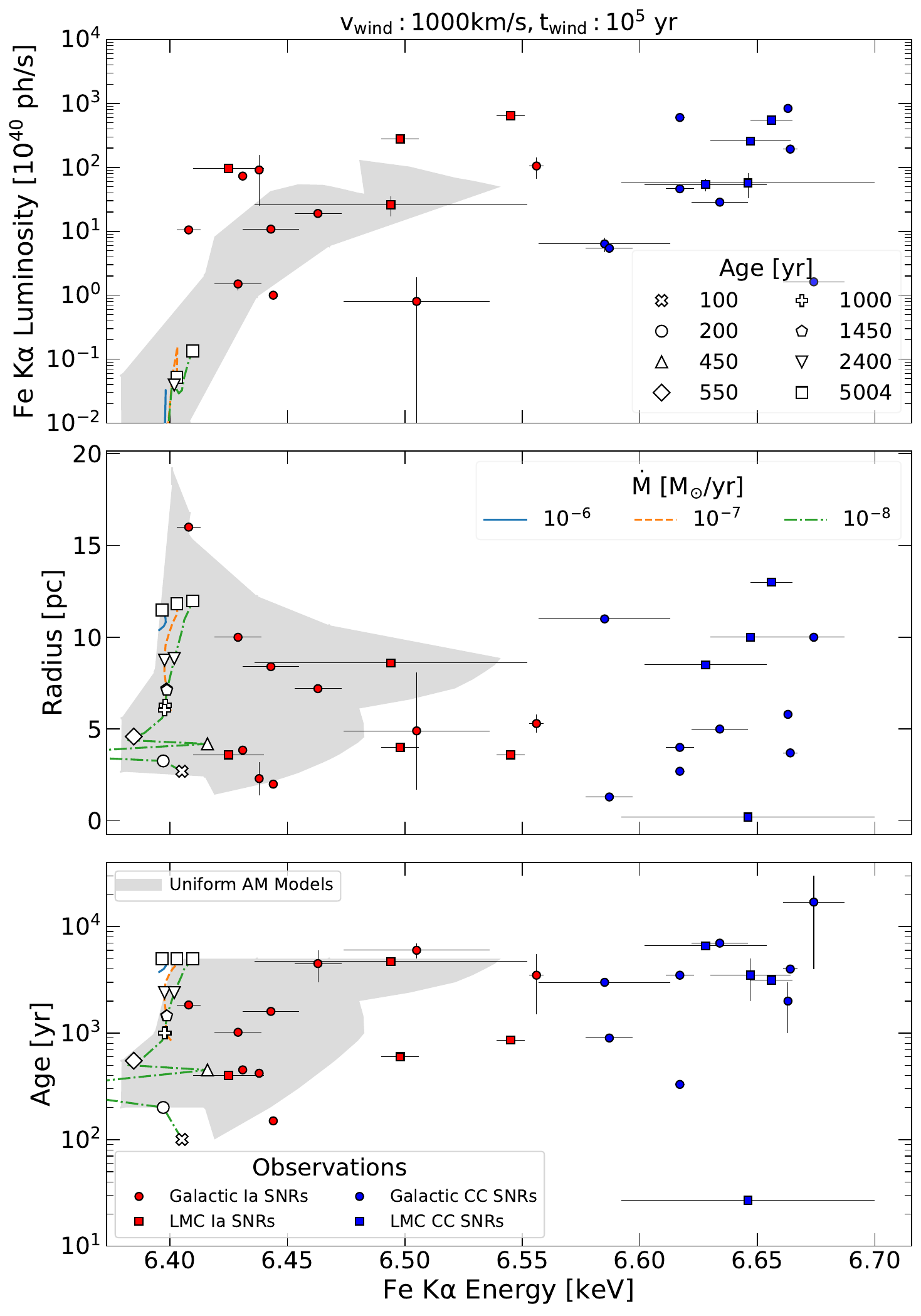}
    \caption{Comparison between SNR models interacting with fast outflows and SNR observations: \vwind\ $=100$ km/s (left column) and \vwind\ $=1000$ km/s (right column).  Bulk properties are shown as a function of \feka\ centroid energy: \feka\ luminosity (\textit{top row}), forward shock radius (\textit{middle row}), and remnant age (\textit{bottom row}). The dashed-dotted green line corresponds to $\mathrm{\dot{M}=10^{-8}\ M_{\odot}/yr}$, the dashed orange line corresponds to $\mathrm{\dot{M}=10^{-7}\ M_{\odot}/yr}$, and the solid blue line corresponds to $\mathrm{\dot{M}=10^{-6}\ M_{\odot}/yr}$, with empty symbols indicating specific SNR ages. Observed values for Type Ia SNRs are shown with filled red symbols, while Core Collapse SNRs are shown with filled blue symbols. The shape of these symbols (circles and squares) distinguishes Milky Way from LMC SNRs. A shaded region corresponding to the parameter space spanned by the uniform $\mathrm{\rho_{\mathrm{AM}}}$ models is included for comparison.}
    \label{fig:fast outflows}
\end{figure*}

Similar to the slow outflow models with low \mdot, we see that for fast outflows with \vwind\ $=100$ km/s and \twind\ $=10^5$ yr, most of the interaction with the CSM structure occurs in the first few hundred years after the SN explosion. After 550 years, the ionization state in these models, shown in Figure~\ref{fig:fast outflows}, is again very close to the uniform $\rho_{\mathrm{AM}} = 10^{-24}$ g/cm$^3$ model, because the forward shock has overrun most of the CSM by this time. While outflows with \vwind\ $=100$ km/s are energy driven and do create low density cavities around the progenitor, these cavities are small in size and do not result in large changes to the bulk SNR dynamics when compared to slow outflows. Much like their slow counterparts, the \feka\ luminosities of these models are offset from most Type Ia SNR observations by at least an order of magnitude. For similar reasons, this problem could also be solved by increasing the ISM density.

In contrast, SNR models that interact with the energy-driven CSM structures created by the fastest outflows in our grid (\vwind\ $=1000$ km/s, \twind\ $=10^5$ yr) are very different from the models that interact with a uniform AM (see Figure~\ref{fig:fast outflows}, right plot). The high mechanical luminosities of these outflows carve low-density cavities around the progenitor which are several pc in size, resulting in much lower ionization timescales in SNR models of a given age \citep{Dwarkadas2005}. The impact on the spectral properties (\feka\ centroids and luminosities) is large, with ionization timescales that are too low for most Type Ia SNRs, and \feka\ luminosities that remain below the thermal continuum for thousands of years after the explosion.


In contrast to the behavior seen in slow outflow models, higher values of \mdot\ lead to \textit{lower} ionization timescales in fast outflow models. This is because for energy driven outflows higher mechanical luminosities lead to larger cavities and lower densities (see Figure~\ref{fig:9_panel}), which result in less ionized plasma. 

\subsection{Outflow duration ($\mathrm{t_{wind}}$)}
\label{sec:twind}

In Sections \ref{sec:slow} and \ref{sec:fast} we described the properties of SNR models interacting with slow and fast progenitor outflows generated over a timescale of $10^5$ yr. Here we describe the impact of increasing the outflow timescale to $10^6$ yr.

\begin{figure}
    \centering
    \includegraphics[width=\linewidth]{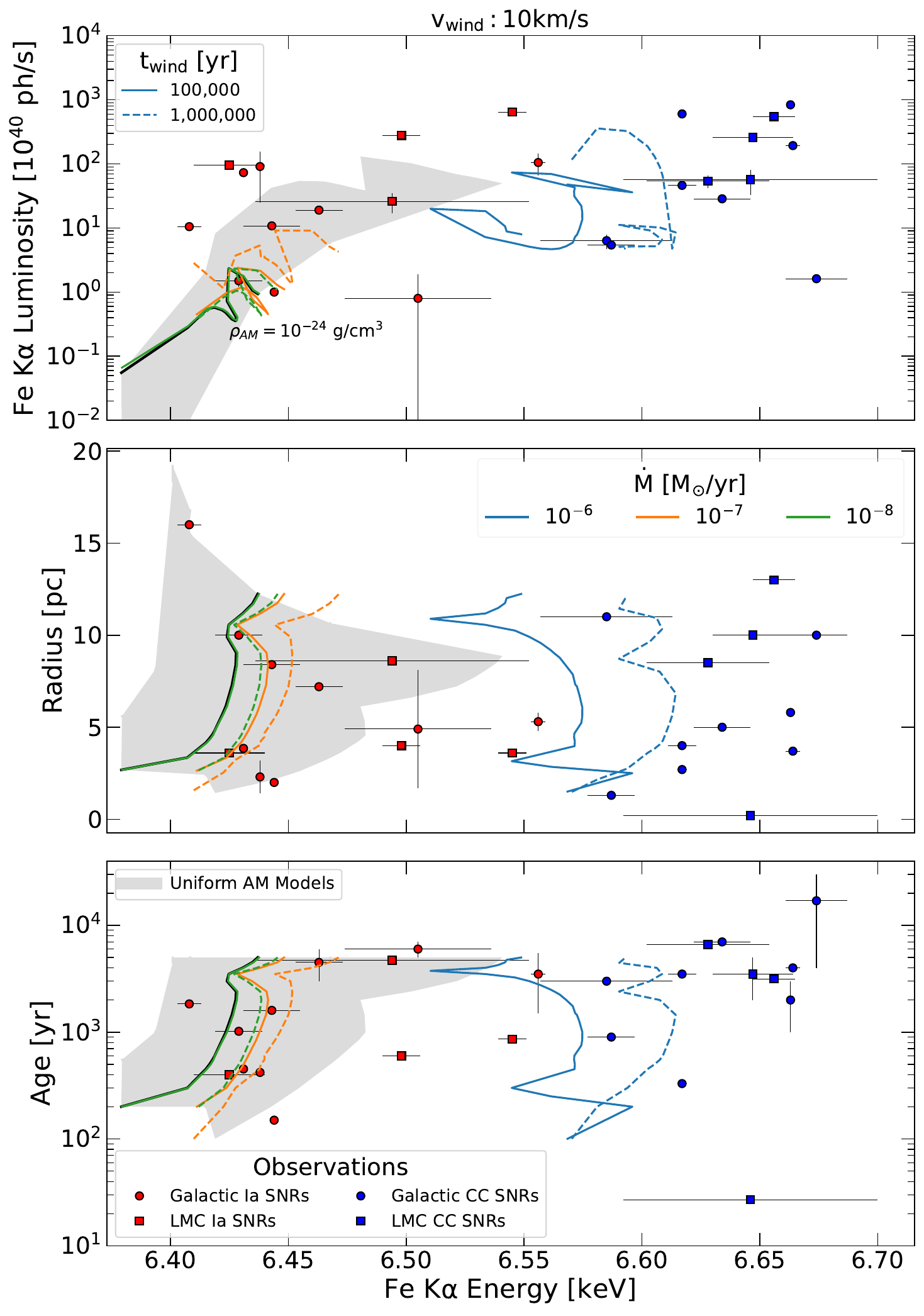}
    \caption{Comparison between \twind\ of $10^5$ and $10^6$ years. Bulk properties are shown as a function of \feka\ centroid energy: \feka\ luminosity (\textit{top row}), forward shock radius (\textit{middle row}), and remnant age (\textit{bottom row}). The solid lines correspond to \twind\ $=10^5$ years and the dotted line corresponds to \twind\ $=10^6$ years. The green lines correspond to $\mathrm{\dot{M}=10^{-8}\ M_{\odot}/yr}$, the orange lines correspond to $\mathrm{\dot{M}=10^{-7}\ M_{\odot}/yr}$, and the blue lines correspond to $\mathrm{\dot{M}=10^{-6}\ M_{\odot}/yr}$. A solid black line is overlaid to represent an SNR interacting with a uniform density of $\mathrm{\rho_{\mathrm{AM}}=1.0\times10^{-24}\ g/cm^3}$. A shaded region corresponding to the parameter space spanned by the uniform $\mathrm{\rho_{\mathrm{AM}}}$ models is included for comparison.}
    \label{fig:twind}
\end{figure}

For slow (\vwind\ $=10$ km/s) outflows, increasing \twind\ to $10^6$ yr makes the momentum-driven CSM structures larger (see Figure~\ref{fig:9_panel}). The effect this has on SNR models expanding into these CSM structures is shown in Figure~\ref{fig:twind}. We see an increase in both \feka\ centroid energy and luminosity for all values of \mdot, with the effect being stronger at higher \mdot. For the model with \mdot\ $=10^{-6}$ \msun/yr, the \feka\ centroid energies increase to 6.56-6.62 keV, well into the CC SNR range. In other words, an isotropic outflow that deposits 1 \msun\ of material within $\sim$ 4 pc of the progenitor before the SN explosion is clearly incompatible with the bulk dynamics of known Type Ia SNRs. At lower values of \mdot, the effect of increasing \twind\ is more modest. The model with \mdot\ $=10^{-7}$ \msun/yr  becomes about an order of magnitude more luminous in \feka, which improves the agreement with observations, but not to the point where it can reproduce most of the SNRs in the sample. The changes to the \mdot\ $=10^{-8}$ \msun/yr are much smaller.



For fast outflows (\vwind\ $\geq 100$ km/s), an increase in \twind\ leads to larger cavities around the progenitor, as shown in Figure~\ref{fig:9_panel}, which results in SNR models with lower \feka\ luminosities and centroids. For models with \vwind\ $=100$ km/s, this removes the small overlap seen with SNR observations in Figure~\ref{fig:fast outflows} at \twind\ $=10^5$ yr (which, recall, was due to the fact that after a certain age the SNR is interacting with the uniform ISM outside the CSM structure). For models with  \vwind\ $=1000$ km/s, the increase in \twind\ results in an even larger difference with the observations.



\subsection{Collisionless Electron Heating at the Reverse Shock ($\beta$)} \label{sec:beta}

\begin{figure}
    \centering
    \includegraphics[width=\columnwidth]{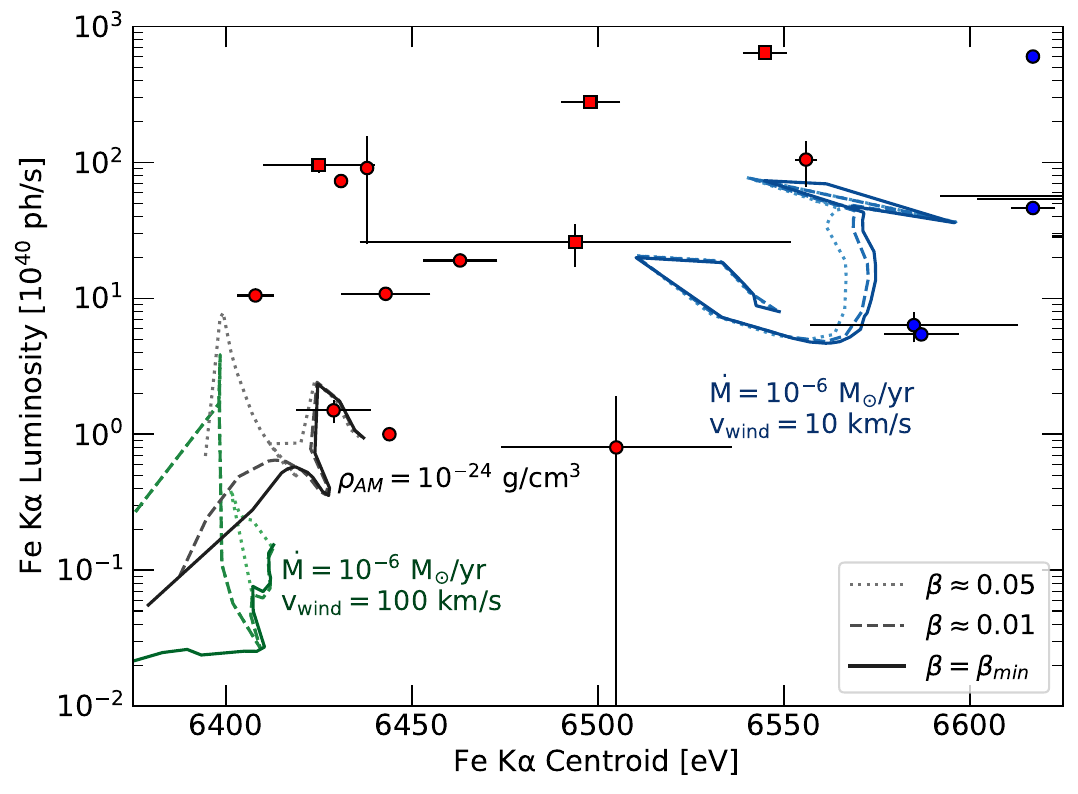}
    \caption{\feka\ luminosities and centroid for SNR models with varying $\beta$. The blue and green lines correspond to outflow models with \mdot\ $=10^{-6}$ \msun/yr and \vwind\ $=10$ and 100 km/s, respectively. The black lines correspond to a uniform AM model with $\rho_{\mathrm{AM}}=10^{-24}$ g/cm$^3$. For each model, solid lines represent $\beta=\beta_{\mathrm{min}}$, dashed lines $\beta=0.01$, and dotted lines $\beta=0.05$. The observations for Type Ia SNRs are shown with red symbols.}
    \label{fig:beta_winds}
\end{figure}

To evaluate the impact of collisionless electron heating at the reverse shock on our spectral calculations,  we calculated SNR models with values of $\beta$ spanning the range preferred by observations: $\beta=\beta_{\mathrm{min}}$, $\beta=0.01$, and $\beta=0.05$ \citep{badenes_thermal_2005,yamaguchi_new_2014}. For this purpose, we choose one uniform AM model ($\rho_{\mathrm{AM}}=1\times10^{-24}$ g/cm$^3$), one slow outflow model (\vwind\ $=10$ km/s, \mdot\ $=10^{-6}$ \msun/yr), and one fast outflow model (\vwind\ $=100$ km/s, \mdot\ $=10^{-6}$ \msun/yr). The results are shown in Figure~\ref{fig:beta_winds}. In agreement with \citealt{badenes_thermal_2005}, we find that increasing $\beta$ can have a large impact on the \feka\ emission in SNR models, but this impact is not uniform across the parameter space. For SNR models evolving in higher densities (such as the slow outflow model we consider here), the impact of $\beta$ is modest, because the additional thermal energy imparted on the electrons at the reverse shock is rapidly diluted by a large influx of colder electrons ejected by the ongoing collisional ionization -- this is sometimes referred to as ionization cooling \citep{yamaguchi_new_2014}. In SNR models interacting with a lower density, like the fast outflow model and the uniform AM model shown in Figure~\ref{fig:beta_winds}, increasing $\beta$ has a more noticeable impact, leading to a modest decrease in the \feka\ centroid and a large increase in the \feka\ luminosity. This is because larger electron temperatures translate into modest reductions in ionization rates, and large increases in plasma emissivities \citep{badenes_thermal_2005}. We note, however, that the increase in \feka\ luminosities is largest at early SNR ages, when the ionization timescale in the shocked ejecta is still too low to match most known Type Ia SNRs.

\section{Discussion} \label{sec:discussion}

\subsection{Comparison to Observations: Uniform AM vs. CSM models}
\label{subsec:CompObs}


The goal of the present study is not to produce viable models for any specific objects, but rather to use HD+NEI models to understand the bulk dynamics of Type Ia SNRs as a class. For this purpose, the most interesting objects are the historical SNRs (Kepler, Tycho, SN 1006, and RCW 86), since their known ages (421, 453, 1019, and 1840 yr -- see \citealt{stephenson_historical_2002} and Table~\ref{tab:TypeIaSNRs}) put the strongest constraints on comparisons to HD+NEI models. To this group we can add the LMC SNRs with light echoes (0509-67.5, 0519-69.0 and N103B), which have independent age estimates ($\sim$400, $\sim$600, and $\sim$860 years, respectively, \citealt{rest_light_2005,rest_spectral_2008}), and SNR G1.9+0.3, whose small radius and rapid expansion rate require an age of $\lesssim$150 yr \citep{reynolds_x-ray_2009,carlton_expansion_2011,sarbadhicary_two_2019,griffeth_stone_type_2021}. The remaining five objects in the compilation of \citealt{martinez-rodriguez_chandrasekhar_2018} (DEM L71, 3C397, G344.7-0.1, G352.7-0.1, and G337.2-0.7) have less reliable age estimates derived using dynamical and spectral arguments, and should be treated with more caution. With the exception of the LMC SNR DEM L71, these are also Galactic objects with somewhat uncertain distances, which can affect their radii and \feka\ luminosity estimates. In the following discussion, we will put an emphasis on the eight SNRs with the best age constraints, and qualify our conclusions regarding the other five SNRs by taking into account the uncertainties in their ages.


\begin{figure*}
    \centering
    \includegraphics[width=\linewidth]{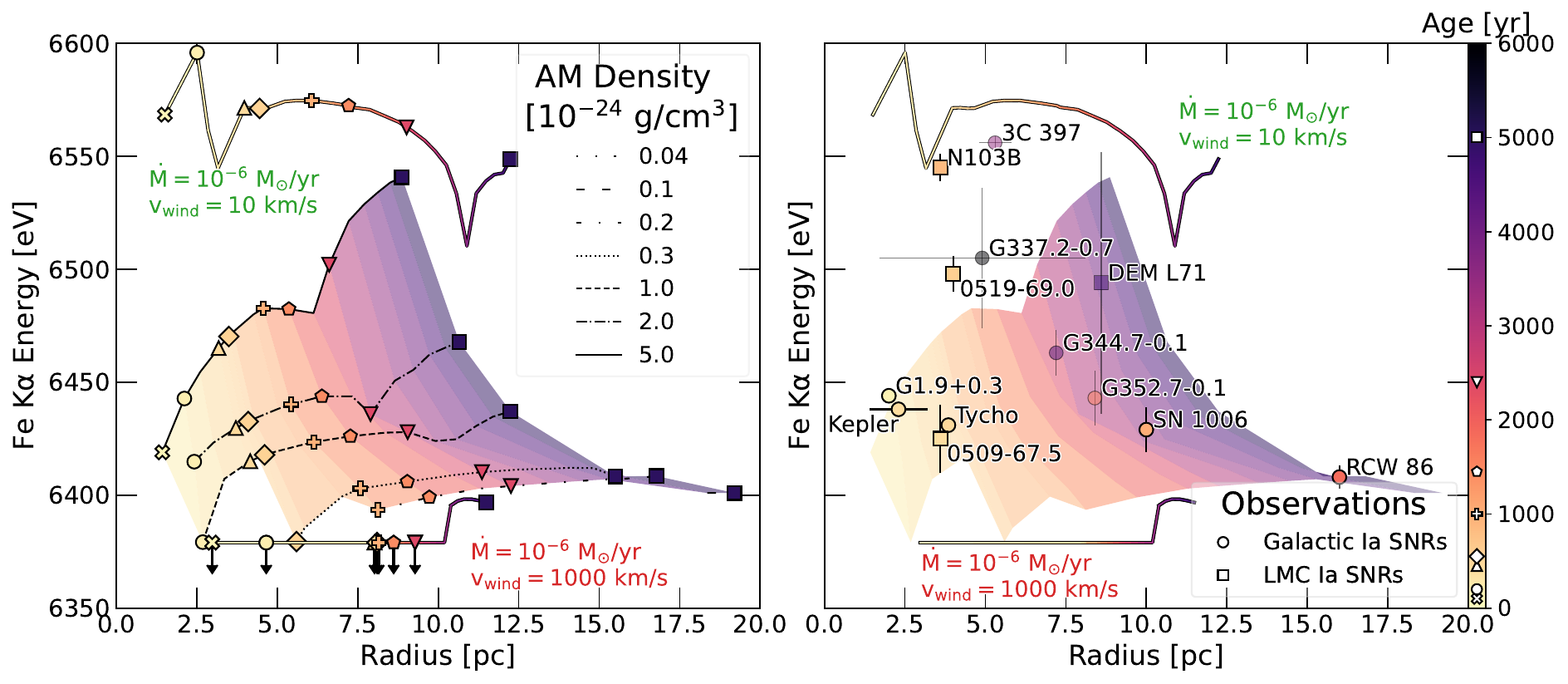}
    \caption{The parameter space of SNR model and observations. (\textit{left}) SNR models are shown with a color gradient corresponding to the age of the model, and symbols corresponding to specific ages, as shown in the color bar to the right. Isotropic outflow models are shown with a \mdot\ of $10^{-6}$ \msun/yr and \vwind\ of 10 and 1000 km/s to bracket the wind velocity parameter space. Upper limits in \feka\ centroid energy are marked with black arrows on the 1000 km/s model as the synthetic spectra do not have a sufficient \feka\ signal to register a centroid energy. (\textit{right}) Observations of Type Ia SNRs are overlain with the same color gradient. Objects with uncertain ages (3C 397, G344.7-0.1, G337.2-0.7, DEM L71, and G352.7-0.1) are plotted with more transparency. }
    \label{fig:legacy}
\end{figure*}

A useful way to visualize the ability of HD+NEI models to reproduce the bulk dynamics of SNRs is presented in Figure~\ref{fig:legacy}. On the left panel of this Figure, we show the radii and \feka\ centroids of our HD+NEI models color coded by SNR age. The uniform AM models form a sequence, with \feka\ centroid and SNR radius increasing with SNR age in each model. Models expanding into lower $\rho_{\mathrm{AM}}$ lead to larger radii and lower \feka\ centroids (towards the bottom), while models expanding into higher $\rho_{\mathrm{AM}}$ lead to smaller radii and higher \feka\ centroids (towards the top). We also show two representative cases of models with CSM interaction. The slow outflow model with \vwind\ $=10$ km/s, \mdot\ $=10^{-6}$ \msun/yr, and \twind\ $=10^5$ yr has a more complex behavior with age, reflecting the interaction between the SN ejecta and the CSM structure shown in Figure~\ref{fig:9_panel}, but predicts higher \feka\ centroids than any uniform AM model. The fast outflow model with \vwind\ $=1000$ km/s, \mdot\ $=10^{-6}$ \msun/yr, and \twind\ $=10^5$ yr only yields upper limits to the \feka\ centroid while the ejecta expand inside the cavity, and even at late ages predicts lower \feka\ centroids than any uniform AM model. On the right panel of this Figure, we overlay on these models the observations of the real SNRs in our sample, also color coded by their ages. This comparison only leaves out the \feka\ luminosity, which we discuss briefly here where relevant, and in further detail in the Appendix. Based on their placement on Figure~\ref{fig:legacy}, we can divide the Type Ia SNRs with well determined ages into three main groups:

\begin{itemize}
 \item \textbf{SNRs compatible with a uniform AM interaction:} G1.9+0.3, 0509-67.5, Kepler, and Tycho can all be matched in radius and \feka\ centroid by uniform AM models of the correct age. For 0509-67.5, Kepler, and Tycho, the one parameter that falls short is the \feka\ luminosity, as shown in Figure~\ref{fig:uniform}. \citealt{badenes_constraints_2006}, \citealt{badenes_persistence_2008}, and \citealt{patnaude_origin_2012} found that the introduction of collisionless electron heating at the reverse shock can increase the \feka\ luminosity in uniform AM models to match the observations in these SNRs (see Figure~\ref{fig:beta_winds} and accompanying discussion). SNR 0509-67.5 and Kepler also require a SN Ia model with a higher $^{56}$Ni yield and kinetic energy than \texttt{ddt24}, see \citealt{badenes_persistence_2008,rest_spectral_2008,patnaude_origin_2012,arunachalam_hydro-based_2022}. For G1.9+0.3, the densest uniform AM model matches the SNR radius and \feka\ centroid well, but overpredicts the \feka\ luminosity by 1.5 orders of magnitude (see Figure~\ref{fig:uniform}). This discrepancy might be solved using a SN Ia model with a lower $^{56}$Ni yield \citep{badenes_thermal_2003, badenes_constraints_2006}.
 \item \textbf{SNRs that are likely cavity explosions:} This group is comprised by SN 1006 and RCW 86, which overlap the locus of uniform AM models, but have radii that are too large for their ages (i.e., the color of the SNR symbol does not mach the background color on the right panel of Figure~\ref{fig:legacy}). Even though distance uncertainties can affect the radius estimates for these Galactic SNRs, in both cases the model that gives the correct \feka\ centroid at the correct age ($\rho_{\mathrm{AM}} = 10^{-24}$ g/cm$^3$ for SN 1006, $\rho_{\mathrm{AM}} = 0.3\times10^{-24}$ g/cm$^3$ for RCW 86) underpredicts the radius by roughly a factor 2, which is much larger than the distance uncertainties listed in Table~\ref{tab:TypeIaSNRs}. Given their large radii and low \feka\ centroids, these SNRs are good candidates for cavity CSM models, as noted by several authors (see \citealt{badenes_are_2007,williams_rcw_2011,broersen_many_2014} for RCW 86, \citealt{badenes_are_2007,sano_expanding_2022} for SN 1006).
 \item \textbf{SNRs that require dense CSM:} This group is comprised by SNRs 0519-69.0 and N103B. These objects are clearly outside the parameter space covered by uniform AM models in Figure~\ref{fig:legacy}, with ionization timescales that are are too high to be reproduced by any density commonly found in the ISM. These SNRs are good candidates for a strong CSM interaction, as noted previously (see \citealt{court_type_2024,schindelheim_snr_2024} for 0519-69.0, \citealt{Williams2014,Li2017} for N103B).  
\end{itemize}

Classification is more uncertain for SNRs without well determined ages, but given their location in Figure~\ref{fig:legacy}, SNRs G352.7-0.1, G344.7-0.1, and DEM L71 are consistent with a uniform AM interaction, and likely belong to the first group. SNR 3C397 is clearly outside the parameter space for uniform AM interaction, even accounting for uncertainties in its age and distance \citep{leahy_distance_2016}, and likely belongs to the third group of SNRs with strong CSM interaction. SNR G337.2-0.7 might also belong to the group of SNRs with strong CSM interaction, but the distance and age estimates for this SNR are too uncertain to draw a definite conclusion \citep[see][]{rakowski_can_2006,takata_x-ray_2016}. We summarize the grouping of the SNRs in our sample in Table~\ref{tab:TypeIaSNRsGrouping}.

\begin{table*}
    \centering
    \caption{Grouping of Type Ia SNRs according to their bulk dynamics}
    \begin{tabular}{l c c}
    \hline \hline
    Group & SNRs with independent age estimates & SNRs without independent age estimates \\    
    \hline
    Compatible with uniform AM & G1.9+0.3, 0509-67.5, Kepler, Tycho &  G352.7-0.1, G344.7-0.1, DEM L71. \\
    Cavity Explosions & SN 1006, RCW 86 & \\
    Dense CSM Interaction & 0519-69.0, N103B & 3C397, G337.2-0.7\\
    \hline
    \end{tabular}
    \label{tab:TypeIaSNRsGrouping}
\end{table*}


The properties of our HD+NEI models can be used to provide a physical framework for this grouping. The SNRs in the first group (G1.9+0.3, 0509-67.5, Kepler, and Tycho, plus possibly G352.7-0.1, G344.7-0.1, and DEM L71), had progenitors that did not substantially modify their environment on $\sim$pc scales. Of course, this does not imply that these objects did not interact with some sort of CSM. As we have seen, some of our models (e.g. \mdot\ $\lesssim10^{-7}$ \msun/yr with \vwind\ $\lesssim$100 km/s and \twind=$10^5$ yr) result in small CSM structures with relatively low densities that are overrun by the forward shock in the first few hundred years after the explosion, leaving little or no imprint on the bulk dynamics of the SNR. Under these circumstances, a CSM interaction cannot be ruled out for any specific object, although Occam's razor would always lead us to prefer a uniform AM model. The Kepler SNR is a particularly interesting case. Although the bulk dynamics of this SNR can be explained with a uniform AM interaction, several lines of evidence suggest that Kepler is interacting with some sort of CSM \citep{Blair2007,reynolds_deep_2007}. Several authors have suggested a slow progenitor outflow with a high mass loss rate \citep[$\gtrsim10^{-6}$ \msun/yr][]{chiotellis_imprint_2012,katsuda_keplers_2015}, but this is clearly inconsistent with the properties of the \feka\ emission. Our slow outflow model with \vwind\ $=10$ km/s and \mdot\ $=10^{-6}$ \msun/yr does match the \feka\ luminosity of Kepler, but grossly overpredicts the \feka\ centroid, as shown in Figure~\ref{fig:legacy}. More complex outflow models have been shown to work for Kepler using HD+NEI simulations \citep[see][and the discussion in Section~\ref{subsec:outside} below]{patnaude_origin_2012}. Similar scenarios might apply to other objects in this group, but it is clear that these SNRs are not compatible with either slow isotropic outflows that leave behind large amounts of dense CSM close to the progenitor or fast outflows with high mechanical luminosities that excavate large low-density cavities, as seen on Figure~\ref{fig:legacy}. This implies that structures like the 13$\times$27 pc 'ring' reported by \cite{chen_large-field_2017} around Tycho are either not associated with the progenitor, or not due to canonical fast outflows with large mechanical luminosities like the ones we explore here.

Although the SNRs in the second group (SN 1006 and RCW 86) are too large to be explained by uniform AM interaction, the fast outflow models in our grid do not provide a satisfactory approximation to their bulk dynamics. Our fastest outflows (\vwind\ $=1000$ km/s) do excavate large cavities ($\gtrsim$10 pc - see Figure~\ref{fig:9_panel}), but the densities inside these cavities are so low that the \feka\ centroids fall short, and the \feka\ luminosities are orders of magnitude below the observations. The case of RCW 86 is particularly interesting, since the size, expansion parameters, and spectral properties of this SNR are broadly consistent with an SN Ia explosion in a large low-density cavity, presumably excavated by a fast, sustained outflow from the progenitor \citep{vink_x-ray_2006,badenes_are_2007,williams_rcw_2011,broersen_many_2014}. The poor match between RCW 86 and our fast outflow models suggests that the properties of the progenitor outflow in this SNR must have been very different from the values we explore here, or must have had large deviations from our assumptions of spatial isotropy or temporal invariance. 

The SNRs in the third group (0519-69.0, N103B, 3C 397, and possibly G337.2-0.7) require some sort of strong CSM interaction. The slow outflow model shown on Figure~\ref{fig:legacy} gives an acceptable approximation to the radius and \feka\ centroid of N103B at the SNR age, though the \feka\ luminosity is still an order of magnitude too low. These discrepancies are small enough to be bridged by an increase in collisionless electron heating, a more energetic SN Ia model, or slightly different values of \mdot\ or \twind. For the other three objects in this group, our simple slow isotropic outflows do not seem to work, but more complex mass loss histories might \citep[e.g.][, see Section~\ref{subsec:outside}]{schindelheim_snr_2024}.

\subsection{CSM interaction outside our model grid}
\label{subsec:outside}

While we have made an effort to produce a comprehensive grid for CSM interaction in Type Ia SNRs, the parameter space for progenitor outflows is large, and there might be many promising models outside our grid. Again, the Kepler SNR is an interesting case within the first group of objects. The only work to perform full HD+NEI calculations for Kepler assuming a Type Ia origin \citep{patnaude_origin_2012} found that a slow outflow with \mdot\ ($6\times10^{-6}$ \msun/yr) and \vwind\ (20 km/s) does match the X-ray spectrum and bulk dynamics, but only after carving a small ($\sim$0.03 pc) cavity in the inner part of the outflow, and using a SN Ia model more energetic than \texttt{ddt24}. It is possible that small cavities like these might help to reconcile other SNRs in this group with a CSM
interaction, perhaps produced by mass conservative episodes or short lived fast progenitor outflows just before the explosion. 

The Type Ia SNRs with large radii and low ionization timescales (SN 1006 and RCW 86) are likely associated with low-density cavities excavated by fast progenitor outflows, but those cavities must be quite different from the ones we consider here. One possibility would be to inject fast progenitor outflows into a denser ISM to increase the \feka\ luminosities without making the \feka\ centroids too large for these objects. \cite{badenes_are_2007} found a reasonable match to the radius, shock velocity, and ionization timescale of Si in SN 1006 with a cavity excavated using a time varying progenitor outflow (their model HP3, taken from Figure 1e in \citealt{han_single-degenerate_2004}) with \vwind\ $ = 2000$ km/s, peak \mdot\ $=3\times10^{-7}$ \msun/yr, and \twind\ $=2\times10^{6}$ yr. However, these authors did not comment on the ability of this model to match the \feka\ emission in SN 1006, which was not detected until 2008 \citep{yamaguchi_x-ray_2008}. Our closest outflow model (\vwind\ $=1000$ km/s, \mdot\ $=10^{-7}$ \msun/yr) grossly under-predicts the radius, \feka\ centroid, and \feka\ luminosity at the age of SN 1006.   

As for the SNRs that require a denser CSM, an interesting possibility was explored by \cite{court_type_2024}, who simulated the interaction between SN Ia ejecta and post-common envelope cocoon models from \cite{garcia-segura_common_2018}, as proposed by \cite{kashi_circumbinary_2011}. These cocoons have a complex bipolar structure that cannot be captured by our simple isotropic outflow models. In most cases, as noted by \cite{court_type_2024}, the CSM density around the progenitor is too high, leading to recombining plasmas and \feka\ centroids over 6.65 keV, close to the highest values observed in CC SNRs. However, \cite{schindelheim_snr_2024} were able to match the bulk dynamics of SNR 0519-69.0 using one of these cocoons with less dense CSM, where the material ejected by the common envelope episode expanded for $10^4$ yr before the SN explosion. It is possible that similar models might be able to reproduce the bulk dynamics of 3C397 and G337.2-0.7 as well.

Of course, since our CSM and SNR models are 1D, there are inherent limitations to our methodology. While spherical symmetry is an acceptable first approximation to explore the large parameter space of CSM interaction in Type Ia SNRs and compare to spatially integrated observations, as we have done here, all the objects in our sample are spatially resolved by modern X-ray telescopes. Type Ia SNRs as a class are known to be more symmetric than core collapse SNRs \citep{lopez_typing_2009,lopez_using_2011}, but all spatially resolved SNRs, regardless of type, show some degree of deviation from spherical symmetry. For Type Ia SNRs, these deviations range from modest asymmetries in largely spherical objects like Tycho \citep{sato_genus_2019,mandal_measurement_2024} to more clear departures from spherical symmetry like the central belt and north-south asymmetry in the Kepler SNR \citep{reynolds_deep_2007,burkey_x-ray_2013} or the large scale structures in SNR G1.9+0.3 \citep{griffeth_stone_type_2021}. In principle, SNR asymmetries can stem either from the SN explosion or from anisotropies in the surrounding AM, but in several cases (usually the ones with the most detailed observations) it has been shown conclusively that Type Ia SNRs are not expanding into completely homogeneous material \citep{acero_gas_2007,vink_kinematics_2008,williams_azimuthal_2013,Sarbadhicary2025}. It would be interesting to revisit some of the models presented here with multi-D HD+NEI calculations, as done in \citealt{ferrand_supernova_2021}.

Like all modeling efforts, ours has had to set aside some physical processes that are potentially important to the problem at hand. Chief among these is the back reaction of particle acceleration on the hydrodynamics of SNRs. There have been several studies dealing with the impact of this process on the evolution of SNRs \citep[see][and references therein]{Blondin2001, Ellison2007,patnaude_role_2009,patnaude_role_2010,Slane2014}. It is clear that this has a strong impact on the dynamics of the shocked ambient medium, reducing the thermal emission and decreasing the distance between the blast wave and the contact discontinuity \citep{Blondin2001,warren_cosmic-ray_2005}. However, the impact on the dynamics of the shocked ejecta is less clear -- in principle, models that include cosmic ray acceleration could lead to denser ejecta for the same value of $\rho_{\mathrm{AM}}$, which would increase the \feka\ centroids and luminosities \citep{patnaude_role_2009, patnaude_role_2010}. In practice, the reverse shock in the best observed SNRs is traced by hot plasma \citep{yamaguchi_new_2014}, which is hard to explain if its dynamics are strongly affected by cosmic ray acceleration. 


We conclude this discussion with a word of caution. The parameter space for CSM interaction in SNRs is vast, and the comparisons between models and observations that we present here, while informative, are rather crude. The spatially integrated spectral measurements and bulk parameters that we study here cannot possibly capture the complexity of each individual object. We have left aside the kinematics of the shocked ejecta, which can provide a powerful diagnostic for CSM interaction, particularly at the high spectral resolution that is now possible due to XRISM \citep[e.g.][]{Vink2025}. Future studies will explore the relationship between progenitor mass loss and SNR properties in greater detail, but for now we hope that our work showcases the importance of considering CSM interaction scenarios  \textit{globally}, including their effect on the spectral properties of SNRs through HD+NEI calculations.

\section{Conclusions} \label{sec:conclusion}

We have produced the first extensive grid of spectral models for Type Ia SNRs interacting with a CSM produced by uniform isotropic outflows from the SN progenitor. We have systematically varied three outflow parameters: \vwind\ (10, 100, and 1000 km/s), \mdot\ ($10^{-8}$, $10^{-7}$, and $10^{-6}$ \msun/yr) and \twind\ ($10^{5}$ and $10^{6}$ yr), and explored the impact that this variation has on the structure of the CSM and on the bulk properties of the SNRs that interact with them. We have compared the bulk properties (ages, radii and \feka\ line centroids and luminosities) of a sample of 14 Type Ia SNRs in the Milky Way and the Large Magellanic Cloud with the predictions from this model grid. This comparison has led to a division of this sample into three groups, as shown in Figure~\ref{fig:legacy} and Table~\ref{tab:TypeIaSNRsGrouping}. 

We have found that many (perhaps most) Type Ia SNRs did not have progenitors that substantially modified their surroundings on $\sim$pc scales, at least not to the point of affecting the bulk SNR dynamics hundreds or thousands of years after the explosion. In our sample, this group comprises roughly half (7/13) of the objects: G1.9+0.3, 0509-67.5, Kepler, Tycho, G352.7-0.1, G344.7-0.1, and DEM L71. The fact that the bulk dynamics of these SNRs are consistent with a uniform AM interaction does not imply that their progenitors did not lose any mass, but it does put strong constraints on the structure of the CSM at the time of the SN explosion. Specifically, for these objects we can rule out slow isotropic outflows (\vwind\ $\simeq$10 km/s) with high mass loss rates (\mdot\ $\gtrsim10^{-6}$ \msun/yr), which would have deposited large amounts of dense material close to the progenitor, as well as fast outflows (\vwind$\gtrsim$1000 km/s for \twind\ $=10^5$ yr; \vwind$\gtrsim$100 km/s for \twind\ $=10^6$ yr), which would have left behind large low-density cavities. More complex mass loss histories might be able to explain specific objects in this group that have good evidence for some CSM interaction, like the Kepler SNR \citep{patnaude_origin_2012}. 

Roughly 15\% (2/13) of the objects in our sample have the large radii and low ionization timescales that are the hallmark of cavity explosions. The progenitors of these two SNRs (SN 1006 and RCW 86) must have somehow ejected fast ($\sim$1000 km/s), sustained (\twind\ $\gtrsim10^5$ yr) outflows with large mechanical luminosities, perhaps similar to the accretion winds proposed by \cite{hachisu_new_1996}. Our isotropic, continuous outflow models were not able to reproduce the bulk dynamics of these two objects, but models that relax some of our assumptions might. Interestingly, these are the SNRs in our sample with the strongest nonthermal emission, which is associated with cosmic ray acceleration at the forward shock \citep{Koyama1995,Vink2006}. 

And finally, approximately 30\% (4/13) of the SNRs in our sample show evidence for dense CSM on $\sim$pc scales. In our isotropic outflow models, these densities require slow (\vwind\ $\sim$10 km/s) progenitor outflows with \mdot\ $\gtrsim10^{-6}$ \msun/yr. This group includes 0519-67.5, N103B, 3C397, and perhaps also G337.2-0.7. Our slow outflow models provide an acceptable match to the bulk dynamics of N103B, but the other objects in this group might have had more complex mass loss histories, perhaps similar to the post-common envelope cocoons explored by \cite{court_type_2024} and \cite{schindelheim_snr_2024}. It is possible that a small subset of these SN Ia explosions in denser environments might be related to the rare strongly interacting Type Ia SNe studied by \cite{dubay_late-onset_2022}.  

It is interesting to consider these findings about the bulk dynamics of Type Ia SNRs together with the properties of their birth events that have been gleaned by other means. From light echo spectra and detailed studies of the X-ray spectra of the SNRs themselves, we know that Tycho was formed by a normal SN Ia (\citealt{Krause2008}, \citealt{badenes_constraints_2006}), and SNR 0509-67.5 by an overluminous 91T-like SN Ia (\citealt{rest_spectral_2008}, \citealt{badenes_persistence_2008}). 
Recently, \cite{Das2025} found a double-shell structure traced by optical Ca emission in SNR 0509-67.5 that strongly suggests this object was the result of a double detonation in a sub-Chandrasekhar white dwarf. By contrast, SNR 3C397 is noteworthy because of the large amount of neutronized material (stable Ni and Mn) in its shocked ejecta, which is hard to explain with a sub-Chandrasekhar explosion \citep{Yamaguchi2015,Pranav2017}. \cite{Martinez2017} also found evidence for high neutronization in the ejecta of SNRs G337.2-0.7 and N103B from their Ca/S mass ratios, suggesting near-\mch\ WD progenitors for these explosions as well. Remarkably, SNR N103B is surrounded by a substantially younger stellar population than other Type Ia SNRs in the LMC \citep{Badenes2009,maggi_population_2016}, and shows clear evidence for O and Mg enrichment in the CSM \citep{Guest2022}. Recent nebular observations with JWST have shown that at least some normal SN Ia probably have near-\mch\ progenitors like these SNRs \citep{Kwok2025}. 

Putting all this information together leads us to the following conclusions:

\begin{itemize}
    \item Roughly half ($\sim55$ \%) of SN Ia progenitors do not substantially modify their surroundings on $\sim$pc scales. This group leads to SN Ia with varying luminosities within the normal range, and includes at least one likely product of a double detonation explosion in a sub-\mch\ WD.
    \item The other half of SN Ia progenitors do show evidence for modified environments, and can be further divided in two distinct groups.
    \item Approximately 30\% of SN Ia progenitors leave behind enough high density material to affect the bulk dynamics of their SNRs. There is some evidence to suggest that this group might be associated with near-Chandrasekhar mass progenitors produced in younger stellar populations, which likely lose mass as a byproduct of the accretion process. This could be in the form of sustained or intermittent slow outflows ejected from the system over a long period of time, or more rapid common envelope episodes that take place shortly before the SN.
    \item  Approximately 15\% of SN Ia progenitors excavate large low density cavities, likely through an accretion-related process that ejects fast and sustained outflows from the vicinity of the WD.
   
\end{itemize}

\begin{acknowledgments}
We are grateful to the referee for suggestions that helped improve this manuscript, and to Ashley Ruiter, Philipp Podsiadlowski, and Gilles Ferrand for comments. This research was supported in part by the University of Pittsburgh Center for Research Computing through the resources provided. Specifically, this work used the H2P cluster, which is supported by NSF award number OAC-2117681. T.A.C. and C.B. acknowledge support from Chandra Theory grant Nos. TM0-21004X and TM1-22004X, and XRISM Guest Scientist Grant 80NSSC23K0634. S.-H.L. acknowledges support by JSPS Grant-in-aid No. 24K07092 and by WPI, MEXT, Japan. D.J.P. acknowledges support from the {\sl Chandra} X-ray Center, which is operated by the Smithsonian Institution under NASA contract NAS8-03060. E.B. acknowledges partial support
from the Spanish project PID2021-123110NB-100, financed by
MCIN/AEI/10.13039/501100011033/FEDER/UE. 
\end{acknowledgments}
%



\software{
    \texttt{astropy} \citep{astropy_collaboration_astropy_2013,astropy_collaboration_astropy_2018,astropy_collaboration_astropy_2022},
    \texttt{ChN} \citep{ellison_particle_2007,ellison_efficient_2010, lee_generalized_2012,lee_cr-hydro-nei_2013,lee_reverse_2014,lee_modeling_2015,patnaude_role_2010,patnaude_impact_2017,martinez-rodriguez_chandrasekhar_2018,jacovich_grid_2021}, 
    \texttt{Numpy} \citep{harris_array_2020}, 
    \texttt{Matplotlib} \citep{hunter_matplotlib_2007}, 
    \texttt{Pandas} \citep{mckinney_data_2010},
    \texttt{PyAtomDB} (\url{https://atomdb.readthedocs.io/en/master/index.html}),
    \texttt{SciPy} \citep{virtanen_scipy_2020}
    }

\begin{figure}
    \centering
    \includegraphics[width=\columnwidth]{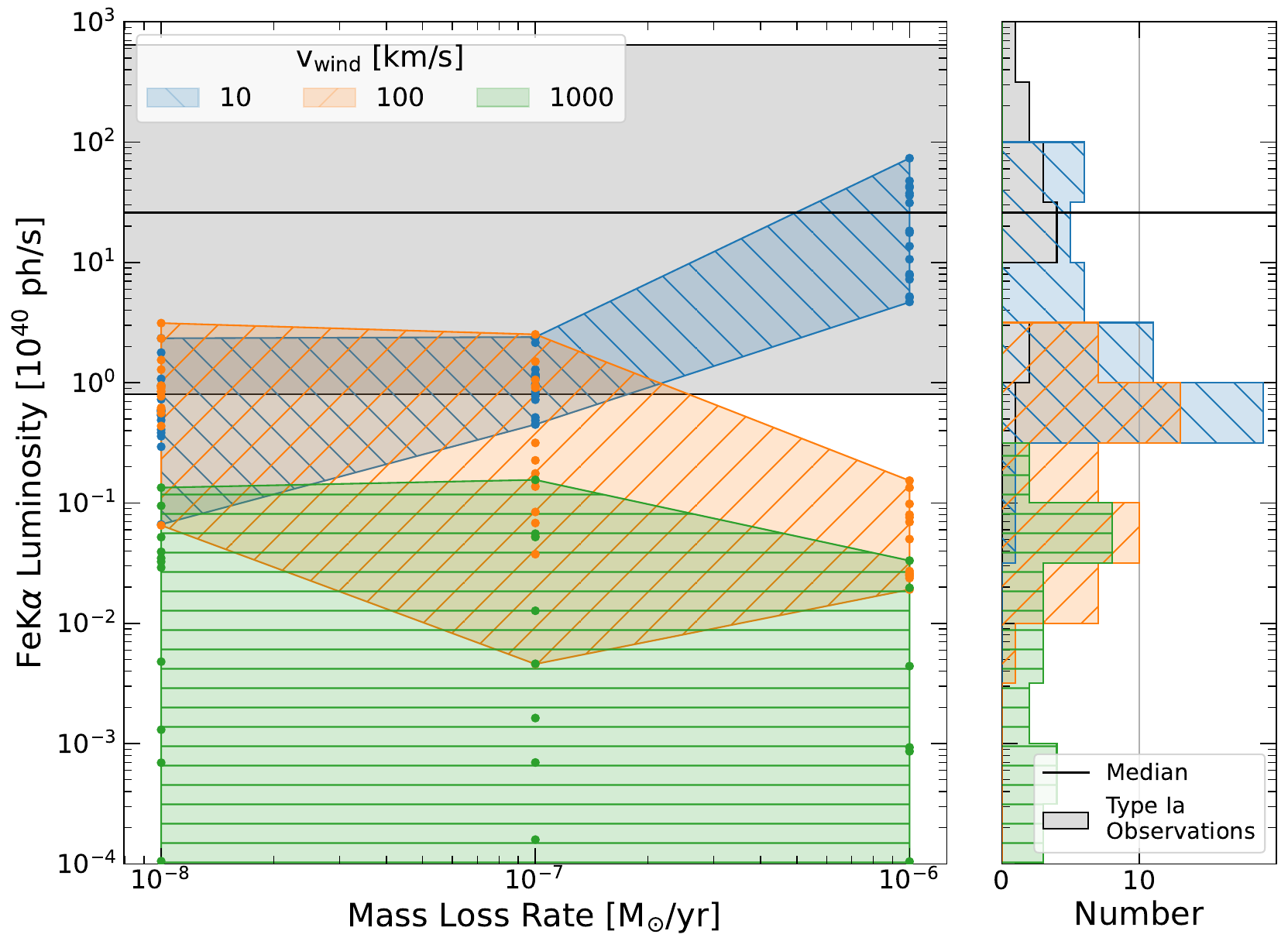}
    \caption{Comparison between the \feka\ luminosity in Models and Observations. (\textit{left}) The \feka\ luminosity is shown as a function of progenitor mass loss rate. The colors indicate the associated \vwind\ and the points correspond to different SNR ages. The shaded regions are the range in values spanned by the models or observations (in grey). 
    (\textit{right}) Histograms of the \feka\ luminosity for each sample. Models with the same \vwind\ are grouped together. The median \feka\ luminosity of the observations is indicated by the horizontal black line. }
    \label{fig:lum_hist_wind}
\end{figure}

\appendix
A common thread in our comparisons between observations and SNR models with CSM interaction has been that the \feka\ luminosity in our models is too low.
To illustrate this point, we show the \feka\ Luminosity as a function of mass loss rate for our CSM models, grouped by \vwind, in
Figure \ref{fig:lum_hist_wind}. This plot showcases the fact that only the slow outflow models with \vwind\ $=10$ km/s have significant overlap with SNR observations, regardless of the value of \mdot. As discussed in Section~\ref{subsec:CSM}, in many slow outflow models the SNR has overrun the CSM and is in fact interacting with the uniform ISM at the relevant SNR ages. The most successful slow outflow models are those with $\dot{M}=10^{-6}$ \msun/yr, which, as we have seen, result in \feka\ centroids that are too high for all Type Ia SNRs except N103B and 3C397. For the fast progenitor outflows, we see that some of the \vwind\ $=100$ km/s models with the lowest \mdot\ values overlap the SNRs with the lowest \feka\ luminosities, but none of the \vwind\ $=1000$ km/s models are luminous enough to reproduce SNR observations at any point in their evolution. This offset between the \feka\ luminosities predicted by CSM interaction models with fast outflows and SNR observations is too large to be resolved by the introduction of collisionless electron heating (Section~\ref{sec:beta}), the use of more energetic SN Ia models, or the mismatch between Eulerian and Lagrangian codes (Section~\ref{sec:uniformAM}).

\begin{figure}
    \centering
    \includegraphics[width=\columnwidth]{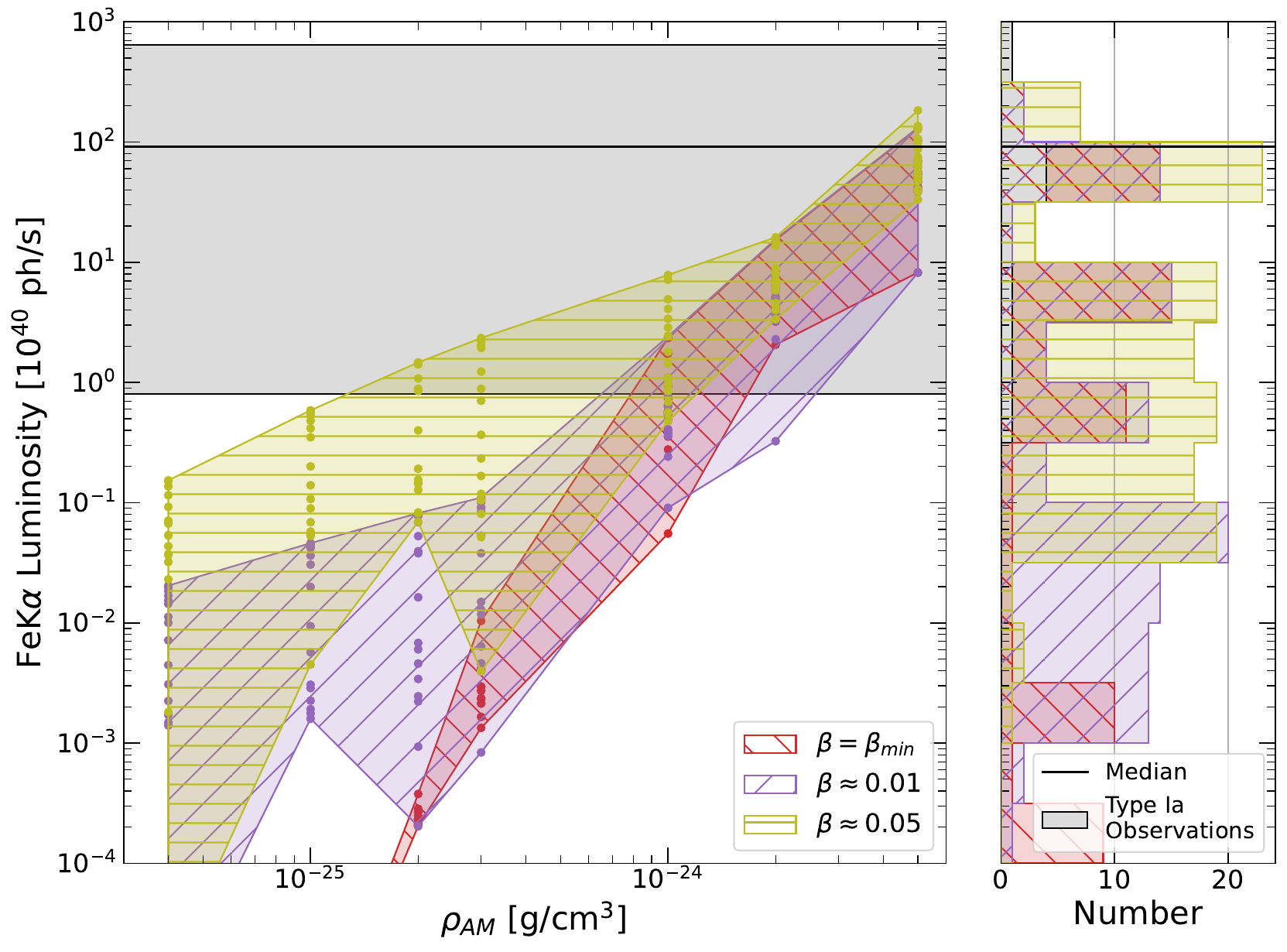}
    \caption{Comparison between the \feka\ luminosity in Models and Observations. (\textit{left}) The \feka\ luminosity is shown as a function of to uniform ambient medium density. The colors indicate the value of $\beta$, and the points correspond to different SNR ages. The shaded regions are the range in values spanned by the models or observations (in grey). 
    (\textit{right}) Histograms of the \feka\ luminosity for each sample. The median luminosity of the observations is indicated by the horizontal black line. }
    \label{fig:beta_uniform}
\end{figure}

An equivalent plot for the uniform AM models is shown in Figure \ref{fig:beta_uniform}. Here, models are shown as a function of $\rho_{\mathrm{AM}}$, grouped by the value of $\beta$. 
Again, the conclusion we can draw from this comparison is that uniform AM models generally perform better than the CSM interaction models, particularly at higher values of $\rho_{\mathrm{AM}}$. This conclusion is strengthened when we consider the other bulk parameters discussed in the previous Section (ages, radii, and \feka\ centroids). We note that the uniform AM densities that provide the better match to the measured \feka\ luminosities ($\rho_{\mathrm{AM}} \gtrsim 10^{-24}\, g\,cm^{-3}$) are noticeably higher than the mean ISM density, which is $\sim0.3\times10^{-24}\, g\,cm^{-3}$ \citep{berkhuijsen_density_2008}.



\bibliography{CSM}{}

\begin{thebibliography}{}
\expandafter\ifx\csname natexlab\endcsname\relax\def\natexlab#1{#1}\fi
\providecommand{\url}[1]{\href{#1}{#1}}
\providecommand{\dodoi}[1]{doi:~\href{http://doi.org/#1}{\nolinkurl{#1}}}
\providecommand{\doeprint}[1]{\href{http://ascl.net/#1}{\nolinkurl{http://ascl.net/#1}}}
\providecommand{\doarXiv}[1]{\href{https://arxiv.org/abs/#1}{\nolinkurl{https://arxiv.org/abs/#1}}}

\bibitem[{Acero {et~al.}(2007)Acero, Ballet, \& Decourchelle}]{acero_gas_2007}
Acero, F., Ballet, J., \& Decourchelle, A. 2007, Astronomy and Astrophysics,
  475, 883, \dodoi{10.1051/0004-6361:20077742}

\bibitem[{Arunachalam {et~al.}(2022)Arunachalam, Hughes, Hovey, \&
  Eriksen}]{arunachalam_hydro-based_2022}
Arunachalam, P., Hughes, J.~P., Hovey, L., \& Eriksen, K. 2022, The
  Astrophysical Journal, 938, 121, \dodoi{10.3847/1538-4357/ac927c}

\bibitem[{{Astropy Collaboration} {et~al.}(2013){Astropy Collaboration},
  Robitaille, Tollerud, Greenfield, Droettboom, Bray, Aldcroft, Davis,
  Ginsburg, Price-Whelan, Kerzendorf, Conley, Crighton, Barbary, Muna,
  Ferguson, Grollier, Parikh, Nair, Unther, Deil, Woillez, Conseil, Kramer,
  Turner, Singer, Fox, Weaver, Zabalza, Edwards, Azalee~Bostroem, Burke, Casey,
  Crawford, Dencheva, Ely, Jenness, Labrie, Lim, Pierfederici, Pontzen, Ptak,
  Refsdal, Servillat, \& Streicher}]{astropy_collaboration_astropy_2013}
{Astropy Collaboration}, Robitaille, T.~P., Tollerud, E.~J., {et~al.} 2013,
  Astronomy and Astrophysics, 558, A33, \dodoi{10.1051/0004-6361/201322068}

\bibitem[{{Astropy Collaboration} {et~al.}(2018){Astropy Collaboration},
  Price-Whelan, Sipőcz, Günther, Lim, Crawford, Conseil, Shupe, Craig,
  Dencheva, Ginsburg, VanderPlas, Bradley, Pérez-Suárez, de~Val-Borro,
  Aldcroft, Cruz, Robitaille, Tollerud, Ardelean, Babej, Bach, Bachetti,
  Bakanov, Bamford, Barentsen, Barmby, Baumbach, Berry, Biscani, Boquien,
  Bostroem, Bouma, Brammer, Bray, Breytenbach, Buddelmeijer, Burke, Calderone,
  Cano~Rodríguez, Cara, Cardoso, Cheedella, Copin, Corrales, Crichton,
  D'Avella, Deil, Depagne, Dietrich, Donath, Droettboom, Earl, Erben, Fabbro,
  Ferreira, Finethy, Fox, Garrison, Gibbons, Goldstein, Gommers, Greco,
  Greenfield, Groener, Grollier, Hagen, Hirst, Homeier, Horton, Hosseinzadeh,
  Hu, Hunkeler, Ivezić, Jain, Jenness, Kanarek, Kendrew, Kern, Kerzendorf,
  Khvalko, King, Kirkby, Kulkarni, Kumar, Lee, Lenz, Littlefair, Ma, Macleod,
  Mastropietro, McCully, Montagnac, Morris, Mueller, Mumford, Muna, Murphy,
  Nelson, Nguyen, Ninan, Nöthe, Ogaz, Oh, Parejko, Parley, Pascual, Patil,
  Patil, Plunkett, Prochaska, Rastogi, Reddy~Janga, Sabater, Sakurikar,
  Seifert, Sherbert, Sherwood-Taylor, Shih, Sick, Silbiger, Singanamalla,
  Singer, Sladen, Sooley, Sornarajah, Streicher, Teuben, Thomas, Tremblay,
  Turner, Terrón, van Kerkwijk, de~la Vega, Watkins, Weaver, Whitmore,
  Woillez, Zabalza, \& {Astropy
  Contributors}}]{astropy_collaboration_astropy_2018}
{Astropy Collaboration}, Price-Whelan, A.~M., Sipőcz, B.~M., {et~al.} 2018,
  The Astronomical Journal, 156, 123, \dodoi{10.3847/1538-3881/aabc4f}

\bibitem[{{Astropy Collaboration} {et~al.}(2022){Astropy Collaboration},
  Price-Whelan, Lim, Earl, Starkman, Bradley, Shupe, Patil, Corrales, Brasseur,
  Nöthe, Donath, Tollerud, Morris, Ginsburg, Vaher, Weaver, Tocknell,
  Jamieson, van Kerkwijk, Robitaille, Merry, Bachetti, Günther, Aldcroft,
  Alvarado-Montes, Archibald, Bódi, Bapat, Barentsen, Bazán, Biswas, Boquien,
  Burke, Cara, Cara, Conroy, Conseil, Craig, Cross, Cruz, D'Eugenio, Dencheva,
  Devillepoix, Dietrich, Eigenbrot, Erben, Ferreira, Foreman-Mackey, Fox,
  Freij, Garg, Geda, Glattly, Gondhalekar, Gordon, Grant, Greenfield, Groener,
  Guest, Gurovich, Handberg, Hart, Hatfield-Dodds, Homeier, Hosseinzadeh,
  Jenness, Jones, Joseph, Kalmbach, Karamehmetoglu, Kałuszyński, Kelley,
  Kern, Kerzendorf, Koch, Kulumani, Lee, Ly, Ma, MacBride, Maljaars, Muna,
  Murphy, Norman, O'Steen, Oman, Pacifici, Pascual, Pascual-Granado, Patil,
  Perren, Pickering, Rastogi, Roulston, Ryan, Rykoff, Sabater, Sakurikar,
  Salgado, Sanghi, Saunders, Savchenko, Schwardt, Seifert-Eckert, Shih, Jain,
  Shukla, Sick, Simpson, Singanamalla, Singer, Singhal, Sinha, Sipőcz,
  Spitler, Stansby, Streicher, Šumak, Swinbank, Taranu, Tewary, Tremblay,
  de~Val-Borro, Van~Kooten, Vasović, Verma, de~Miranda~Cardoso, Williams,
  Wilson, Winkel, Wood-Vasey, Xue, Yoachim, Zhang, Zonca, \& {Astropy Project
  Contributors}}]{astropy_collaboration_astropy_2022}
{Astropy Collaboration}, Price-Whelan, A.~M., Lim, P.~L., {et~al.} 2022, The
  Astrophysical Journal, 935, 167, \dodoi{10.3847/1538-4357/ac7c74}

\bibitem[{Badenes {et~al.}(2005)Badenes, Borkowski, \&
  Bravo}]{badenes_thermal_2005}
Badenes, C., Borkowski, K.~J., \& Bravo, E. 2005, The Astrophysical Journal,
  624, 198, \dodoi{10.1086/428829}

\bibitem[{Badenes {et~al.}(2006)Badenes, Borkowski, Hughes, Hwang, \&
  Bravo}]{badenes_constraints_2006}
Badenes, C., Borkowski, K.~J., Hughes, J.~P., Hwang, U., \& Bravo, E. 2006, The
  Astrophysical Journal, 645, 1373, \dodoi{10.1086/504399}

\bibitem[{Badenes {et~al.}(2003)Badenes, Bravo, Borkowski, \&
  Domínguez}]{badenes_thermal_2003}
Badenes, C., Bravo, E., Borkowski, K.~J., \& Domínguez, I. 2003, The
  Astrophysical Journal, 593, 358, \dodoi{10.1086/376448}

\bibitem[{{Badenes} {et~al.}(2009){Badenes}, {Harris}, {Zaritsky}, \&
  {Prieto}}]{Badenes2009}
{Badenes}, C., {Harris}, J., {Zaritsky}, D., \& {Prieto}, J.~L. 2009, \apj,
  700, 727, \dodoi{10.1088/0004-637X/700/1/727}

\bibitem[{Badenes {et~al.}(2007)Badenes, Hughes, Bravo, \&
  Langer}]{badenes_are_2007}
Badenes, C., Hughes, J.~P., Bravo, E., \& Langer, N. 2007, The Astrophysical
  Journal, 662, 472, \dodoi{10.1086/518022}

\bibitem[{Badenes {et~al.}(2008)Badenes, Hughes, Cassam-Chenaï, \&
  Bravo}]{badenes_persistence_2008}
Badenes, C., Hughes, J.~P., Cassam-Chenaï, G., \& Bravo, E. 2008, The
  Astrophysical Journal, 680, 1149, \dodoi{10.1086/524700}

\bibitem[{Berkhuijsen \& Fletcher(2008)}]{berkhuijsen_density_2008}
Berkhuijsen, E.~M., \& Fletcher, A. 2008, Monthly Notices of the Royal
  Astronomical Society, 390, L19, \dodoi{10.1111/j.1745-3933.2008.00526.x}

\bibitem[{{Blair} {et~al.}(2007){Blair}, {Ghavamian}, {Long}, {Williams},
  {Borkowski}, {Reynolds}, \& {Sankrit}}]{Blair2007}
{Blair}, W.~P., {Ghavamian}, P., {Long}, K.~S., {et~al.} 2007, \apj, 662, 998,
  \dodoi{10.1086/518414}

\bibitem[{Blondin {et~al.}(2001)Blondin, Chevalier, \&
  Frierson}]{blondin_pulsar_2001}
Blondin, J.~M., Chevalier, R.~A., \& Frierson, D.~M. 2001, The Astrophysical
  Journal, 563, 806, \dodoi{10.1086/324042}

\bibitem[{Blondin \& Ellison(2001)}]{blondin_rayleigh-taylor_2001}
Blondin, J.~M., \& Ellison, D.~C. 2001, The Astrophysical Journal, 560, 244,
  \dodoi{10.1086/322499}

\bibitem[{{Blondin} \& {Ellison}(2001)}]{Blondin2001}
{Blondin}, J.~M., \& {Ellison}, D.~C. 2001, \apj, 560, 244,
  \dodoi{10.1086/322499}

\bibitem[{Booth {et~al.}(2016)Booth, Mohamed, \&
  Podsiadlowski}]{booth_modelling_2016}
Booth, R.~A., Mohamed, S., \& Podsiadlowski, P. 2016, Monthly Notices of the
  Royal Astronomical Society, 457, 822, \dodoi{10.1093/mnras/stw001}

\bibitem[{Borkowski {et~al.}(2014)Borkowski, Reynolds, Green, Hwang, Petre,
  Krishnamurthy, \& Willett}]{borkowski_nonuniform_2014}
Borkowski, K.~J., Reynolds, S.~P., Green, D.~A., {et~al.} 2014, The
  Astrophysical Journal, 790, L18, \dodoi{10.1088/2041-8205/790/2/L18}

\bibitem[{Borkowski {et~al.}(2013)Borkowski, Reynolds, Hwang, Green, Petre,
  Krishnamurthy, \& Willett}]{borkowski_supernova_2013}
Borkowski, K.~J., Reynolds, S.~P., Hwang, U., {et~al.} 2013, The Astrophysical
  Journal, 771, L9, \dodoi{10.1088/2041-8205/771/1/L9}

\bibitem[{Bravo {et~al.}(2019)Bravo, Badenes, \&
  Martínez-Rodríguez}]{bravo_snr-calibrated_2019}
Bravo, E., Badenes, C., \& Martínez-Rodríguez, H. 2019, Monthly Notices of
  the Royal Astronomical Society, 482, 4346, \dodoi{10.1093/mnras/sty2951}

\bibitem[{Broersen {et~al.}(2014)Broersen, Chiotellis, Vink, \&
  Bamba}]{broersen_many_2014}
Broersen, S., Chiotellis, A., Vink, J., \& Bamba, A. 2014, Monthly Notices of
  the Royal Astronomical Society, 441, 3040, \dodoi{10.1093/mnras/stu667}

\bibitem[{Burkey {et~al.}(2013)Burkey, Reynolds, Borkowski, \&
  Blondin}]{burkey_x-ray_2013}
Burkey, M.~T., Reynolds, S.~P., Borkowski, K.~J., \& Blondin, J.~M. 2013, The
  Astrophysical Journal, 764, 63, \dodoi{10.1088/0004-637X/764/1/63}

\bibitem[{Carlton {et~al.}(2011)Carlton, Borkowski, Reynolds, Hwang, Petre,
  Green, Krishnamurthy, \& Willett}]{carlton_expansion_2011}
Carlton, A.~K., Borkowski, K.~J., Reynolds, S.~P., {et~al.} 2011, The
  Astrophysical Journal, 737, L22, \dodoi{10.1088/2041-8205/737/1/L22}

\bibitem[{Castor {et~al.}(1975)Castor, McCray, \&
  Weaver}]{castor_interstellar_1975}
Castor, J., McCray, R., \& Weaver, R. 1975, The Astrophysical Journal, 200,
  L107, \dodoi{10.1086/181908}

\bibitem[{Cendes {et~al.}(2020)Cendes, Drout, Chomiuk, \&
  Sarbadhicary}]{cendes_thirty_2020}
Cendes, Y., Drout, M.~R., Chomiuk, L., \& Sarbadhicary, S.~K. 2020, The
  Astrophysical Journal, 894, 39, \dodoi{10.3847/1538-4357/ab6b2a}

\bibitem[{Chen {et~al.}(2011)Chen, Han, \& Tout}]{chen_tidally_2011}
Chen, X., Han, Z., \& Tout, C.~A. 2011, The Astrophysical Journal, 735, L31,
  \dodoi{10.1088/2041-8205/735/2/L31}

\bibitem[{Chen {et~al.}(2017)Chen, Xiong, \& Yang}]{chen_large-field_2017}
Chen, X., Xiong, F., \& Yang, J. 2017, Astronomy and Astrophysics, 604, A13,
  \dodoi{10.1051/0004-6361/201630003}

\bibitem[{Chiotellis {et~al.}(2012)Chiotellis, Schure, \&
  Vink}]{chiotellis_imprint_2012}
Chiotellis, A., Schure, K.~M., \& Vink, J. 2012, Astronomy and Astrophysics,
  537, A139, \dodoi{10.1051/0004-6361/201014754}

\bibitem[{Chomiuk {et~al.}(2012)Chomiuk, Soderberg, Moe, Chevalier, Rupen,
  Badenes, Margutti, Fransson, Fong, \& Dittmann}]{chomiuk_evla_2012}
Chomiuk, L., Soderberg, A.~M., Moe, M., {et~al.} 2012, The Astrophysical
  Journal, 750, 164, \dodoi{10.1088/0004-637X/750/2/164}

\bibitem[{Chomiuk {et~al.}(2016)Chomiuk, Soderberg, Chevalier, Bruzewski,
  Foley, Parrent, Strader, Badenes, Fransson, Kamble, Margutti, Rupen, \&
  Simon}]{chomiuk_deep_2016}
Chomiuk, L., Soderberg, A.~M., Chevalier, R.~A., {et~al.} 2016, The
  Astrophysical Journal, 821, 119, \dodoi{10.3847/0004-637X/821/2/119}

\bibitem[{Court {et~al.}(2024)Court, Badenes, Lee, Patnaude, García-Segura, \&
  Bravo}]{court_type_2024}
Court, T., Badenes, C., Lee, S.-H., {et~al.} 2024, The Astrophysical Journal,
  962, 63, \dodoi{10.3847/1538-4357/ad165f}

\bibitem[{{C{\'u}neo} {et~al.}(2023){C{\'u}neo}, {Mu{\~n}oz-Darias},
  {Jim{\'e}nez-Ibarra}, {Panizo-Espinar}, {S{\'a}nchez-Sierras}, {Armas
  Padilla}, {Casares}, {Mata S{\'a}nchez}, {Torres}, {Vincentelli}, \&
  {Ambrifi}}]{Cuneo2023}
{C{\'u}neo}, V.~A., {Mu{\~n}oz-Darias}, T., {Jim{\'e}nez-Ibarra}, F., {et~al.}
  2023, \aap, 679, A85, \dodoi{10.1051/0004-6361/202347265}

\bibitem[{Dang {et~al.}(2024)Dang, Zhou, Sun, Mao, Vink, Zhang, \&
  Domček}]{dang_typing_2024}
Dang, L.-X., Zhou, P., Sun, L., {et~al.} 2024, Monthly Notices of the Royal
  Astronomical Society, 529, 4117, \dodoi{10.1093/mnras/stae805}

\bibitem[{{Das} {et~al.}(2025){Das}, {Seitenzahl}, {Ruiter}, {R{\"o}pke},
  {Pakmor}, {Vogt}, {Collins}, {Ghavamian}, {Sim}, {Williams}, {Taubenberger},
  {Laming}, {Suherli}, {Sutherland}, \& {Rodr{\'\i}guez-Segovia}}]{Das2025}
{Das}, P., {Seitenzahl}, I.~R., {Ruiter}, A.~J., {et~al.} 2025, Nature
  Astronomy, \dodoi{10.1038/s41550-025-02589-5}

\bibitem[{{Dave} {et~al.}(2017){Dave}, {Kashyap}, {Fisher}, {Timmes},
  {Townsley}, \& {Byrohl}}]{Pranav2017}
{Dave}, P., {Kashyap}, R., {Fisher}, R., {et~al.} 2017, \apj, 841, 58,
  \dodoi{10.3847/1538-4357/aa7134}

\bibitem[{Decourchelle {et~al.}(2000)Decourchelle, Ellison, \&
  Ballet}]{decourchelle_thermal_2000}
Decourchelle, A., Ellison, D.~C., \& Ballet, J. 2000, The Astrophysical
  Journal, 543, L57, \dodoi{10.1086/318167}

\bibitem[{{Diaz} {et~al.}(2010){Diaz}, {Williams}, {Luna}, {Moraes}, \&
  {Takeda}}]{Diaz2010}
{Diaz}, M.~P., {Williams}, R.~E., {Luna}, G.~J., {Moraes}, M., \& {Takeda}, L.
  2010, \aj, 140, 1860, \dodoi{10.1088/0004-6256/140/6/1860}

\bibitem[{Dubay {et~al.}(2022)Dubay, Tucker, Do, Shappee, \&
  Anand}]{dubay_late-onset_2022}
Dubay, L.~O., Tucker, M.~A., Do, A., Shappee, B.~J., \& Anand, G.~S. 2022, The
  Astrophysical Journal, 926, 98, \dodoi{10.3847/1538-4357/ac3bb4}

\bibitem[{{Dwarkadas}(2005)}]{Dwarkadas2005}
{Dwarkadas}, V.~V. 2005, \apj, 630, 892, \dodoi{10.1086/432109}

\bibitem[{Ellison {et~al.}(2007)Ellison, Patnaude, Slane, Blasi, \&
  Gabici}]{ellison_particle_2007}
Ellison, D.~C., Patnaude, D.~J., Slane, P., Blasi, P., \& Gabici, S. 2007, The
  Astrophysical Journal, 661, 879, \dodoi{10.1086/517518}

\bibitem[{{Ellison} {et~al.}(2007){Ellison}, {Patnaude}, {Slane}, {Blasi}, \&
  {Gabici}}]{Ellison2007}
{Ellison}, D.~C., {Patnaude}, D.~J., {Slane}, P., {Blasi}, P., \& {Gabici}, S.
  2007, \apj, 661, 879, \dodoi{10.1086/517518}

\bibitem[{Ellison {et~al.}(2010)Ellison, Patnaude, Slane, \&
  Raymond}]{ellison_efficient_2010}
Ellison, D.~C., Patnaude, D.~J., Slane, P., \& Raymond, J. 2010, The
  Astrophysical Journal, 712, 287, \dodoi{10.1088/0004-637X/712/1/287}

\bibitem[{Ferrand {et~al.}(2021)Ferrand, Warren, Ono, Nagataki, Röpke,
  Seitenzahl, Lach, Iwasaki, \& Sato}]{ferrand_supernova_2021}
Ferrand, G., Warren, D.~C., Ono, M., {et~al.} 2021, The Astrophysical Journal,
  906, 93, \dodoi{10.3847/1538-4357/abc951}

\bibitem[{Ferrière(1998)}]{ferriere_global_1998}
Ferrière, K. 1998, The Astrophysical Journal, 497, 759, \dodoi{10.1086/305469}

\bibitem[{Ferrière(2001)}]{ferriere_interstellar_2001}
Ferrière, K.~M. 2001, Reviews of Modern Physics, 73, 1031,
  \dodoi{10.1103/RevModPhys.73.1031}

\bibitem[{Foster \& Heuer(2020)}]{foster_pyatomdb_2020}
Foster, A.~R., \& Heuer, K. 2020, Atoms, 8, 49, \dodoi{10.3390/atoms8030049}

\bibitem[{Fukushima {et~al.}(2020)Fukushima, Yamaguchi, Slane, Park, Katsuda,
  Sano, Lopez, Plucinsky, Kobayashi, \& Matsushita}]{fukushima_element_2020}
Fukushima, K., Yamaguchi, H., Slane, P.~O., {et~al.} 2020, The Astrophysical
  Journal, 897, 62, \dodoi{10.3847/1538-4357/ab94a6}

\bibitem[{García-Segura {et~al.}(2018)García-Segura, Ricker, \&
  Taam}]{garcia-segura_common_2018}
García-Segura, G., Ricker, P.~M., \& Taam, R.~E. 2018, The Astrophysical
  Journal, 860, 19, \dodoi{10.3847/1538-4357/aac08c}

\bibitem[{Ghavamian {et~al.}(2007)Ghavamian, Laming, \&
  Rakowski}]{ghavamian_physical_2007}
Ghavamian, P., Laming, J.~M., \& Rakowski, C.~E. 2007, The Astrophysical
  Journal, 654, L69, \dodoi{10.1086/510740}

\bibitem[{Gnat \& Sternberg(2007)}]{gnat_time-dependent_2007}
Gnat, O., \& Sternberg, A. 2007, The Astrophysical Journal Supplement Series,
  168, 213, \dodoi{10.1086/509786}

\bibitem[{Griffeth~Stone {et~al.}(2021)Griffeth~Stone, Johnson, Blondin,
  Watson, Borkowski, Fröhlich, Seitenzahl, \&
  Reynolds}]{griffeth_stone_type_2021}
Griffeth~Stone, A., Johnson, H.~T., Blondin, J.~M., {et~al.} 2021, The
  Astrophysical Journal, 923, 233, \dodoi{10.3847/1538-4357/ac300f}

\bibitem[{{Guest} {et~al.}(2022){Guest}, {Blair}, {Borkowski}, {Ghavamian},
  {Hendrick}, {Long}, {Petre}, {Raymond}, {Rest}, {Sankrit}, {Seitenzahl}, \&
  {Williams}}]{Guest2022}
{Guest}, B.~T., {Blair}, W.~P., {Borkowski}, K.~J., {et~al.} 2022, \apj, 926,
  207, \dodoi{10.3847/1538-4357/ac4913}

\bibitem[{Hachisu {et~al.}(1996)Hachisu, Kato, \& Nomoto}]{hachisu_new_1996}
Hachisu, I., Kato, M., \& Nomoto, K. 1996, The Astrophysical Journal, 470, L97,
  \dodoi{10.1086/310303}

\bibitem[{Han \& Podsiadlowski(2004)}]{han_single-degenerate_2004}
Han, Z., \& Podsiadlowski, P. 2004, Monthly Notices of the Royal Astronomical
  Society, 350, 1301, \dodoi{10.1111/j.1365-2966.2004.07713.x}

\bibitem[{Harris {et~al.}(2020)Harris, Millman, Van Der~Walt, Gommers,
  Virtanen, Cournapeau, Wieser, Taylor, Berg, Smith, Kern, Picus, Hoyer,
  Van~Kerkwijk, Brett, Haldane, Del~Río, Wiebe, Peterson, Gérard-Marchant,
  Sheppard, Reddy, Weckesser, Abbasi, Gohlke, \& Oliphant}]{harris_array_2020}
Harris, C.~R., Millman, K.~J., Van Der~Walt, S.~J., {et~al.} 2020, Nature, 585,
  357, \dodoi{10.1038/s41586-020-2649-2}

\bibitem[{Helder {et~al.}(2013)Helder, Vink, Bamba, Bleeker, Burrows,
  Ghavamian, \& Yamazaki}]{helder_proper_2013}
Helder, E.~A., Vink, J., Bamba, A., {et~al.} 2013, Monthly Notices of the Royal
  Astronomical Society, 435, 910, \dodoi{10.1093/mnras/stt993}

\bibitem[{Hughes {et~al.}(2003)Hughes, Ghavamian, Rakowski, \&
  Slane}]{hughes_iron-rich_2003}
Hughes, J.~P., Ghavamian, P., Rakowski, C.~E., \& Slane, P.~O. 2003, The
  Astrophysical Journal, 582, L95, \dodoi{10.1086/367760}

\bibitem[{Hunter(2007)}]{hunter_matplotlib_2007}
Hunter, J.~D. 2007, Comput. Sci. Eng., 9, 90, \dodoi{10.1109/MCSE.2007.55}

\bibitem[{Ivanova {et~al.}(2013)Ivanova, Justham, Chen, De~Marco, Fryer,
  Gaburov, Ge, Glebbeek, Han, Li, Lu, Marsh, Podsiadlowski, Potter, Soker,
  Taam, Tauris, van~den Heuvel, \& Webbink}]{ivanova_common_2013}
Ivanova, N., Justham, S., Chen, X., {et~al.} 2013, Astron Astrophys Rev, 21,
  59, \dodoi{10.1007/s00159-013-0059-2}

\bibitem[{Iłkiewicz {et~al.}(2019)Iłkiewicz, Mikołajewska, Belczyński,
  Wiktorowicz, \& Karczmarek}]{ilkiewicz_wind_2019}
Iłkiewicz, K., Mikołajewska, J., Belczyński, K., Wiktorowicz, G., \&
  Karczmarek, P. 2019, Monthly Notices of the Royal Astronomical Society, 485,
  5468, \dodoi{10.1093/mnras/stz760}

\bibitem[{Jacovich {et~al.}(2021)Jacovich, Patnaude, Slane, Badenes, Lee,
  Nagataki, \& Milisavljevic}]{jacovich_grid_2021}
Jacovich, T., Patnaude, D., Slane, P., {et~al.} 2021, The Astrophysical
  Journal, 914, 41, \dodoi{10.3847/1538-4357/abf935}

\bibitem[{Kashi \& Soker(2011)}]{kashi_circumbinary_2011}
Kashi, A., \& Soker, N. 2011, Monthly Notices of the Royal Astronomical
  Society, 417, 1466, \dodoi{10.1111/j.1365-2966.2011.19361.x}

\bibitem[{Katsuda {et~al.}(2015)Katsuda, Mori, Maeda, Tanaka, Koyama, Tsunemi,
  Nakajima, Maeda, Ozaki, \& Petre}]{katsuda_keplers_2015}
Katsuda, S., Mori, K., Maeda, K., {et~al.} 2015, The Astrophysical Journal,
  808, 49, \dodoi{10.1088/0004-637X/808/1/49}

\bibitem[{Khokhlov(1991)}]{khokhlov_delayed_1991}
Khokhlov, A.~M. 1991, Astronomy and Astrophysics, 245, 114.
\newblock \url{https://ui.adsabs.harvard.edu/abs/1991A&A...245..114K}

\bibitem[{Koo \& McKee(1992{\natexlab{a}})}]{koo_dynamics_1992}
Koo, B.-C., \& McKee, C.~F. 1992{\natexlab{a}}, The Astrophysical Journal, 388,
  93, \dodoi{10.1086/171132}

\bibitem[{Koo \& McKee(1992{\natexlab{b}})}]{koo_dynamics_1992-1}
---. 1992{\natexlab{b}}, The Astrophysical Journal, 388, 103,
  \dodoi{10.1086/171133}

\bibitem[{{Koyama} {et~al.}(1995){Koyama}, {Petre}, {Gotthelf}, {Hwang},
  {Matsuura}, {Ozaki}, \& {Holt}}]{Koyama1995}
{Koyama}, K., {Petre}, R., {Gotthelf}, E.~V., {et~al.} 1995, \nat, 378, 255,
  \dodoi{10.1038/378255a0}

\bibitem[{{Krause} {et~al.}(2008){Krause}, {Tanaka}, {Usuda}, {Hattori},
  {Goto}, {Birkmann}, \& {Nomoto}}]{Krause2008}
{Krause}, O., {Tanaka}, M., {Usuda}, T., {et~al.} 2008, \nat, 456, 617,
  \dodoi{10.1038/nature07608}

\bibitem[{{Kwok} {et~al.}(2025){Kwok}, {Liu}, {Jha}, {Blondin}, {Larison},
  {Miller}, {Dai}, {Foley}, {Filippenko}, {Andrews}, {Andrews}, {Auchettl},
  {Badenes}, {Brink}, {Davis}, {Fl{\"o}rs}, {Galbany}, {Graur}, {Howell},
  {Kumar}, {K{\"o}nyves-T{\'o}th}, {LeBaron}, {Macrie}, {Maeda}, {Maguire},
  {McCully}, {Meza-Retamal}, {Padilla Gonzalez}, {Pakmor}, {Pearson}, {Piro},
  {Polin}, {Rehemtulla}, {Rojas-Bravo}, {Sand}, {Sangkachan}, {Schwab},
  {Sears}, {Singh}, {Subrayan}, {Taggart}, {Temim}, {Terwel}, {Tinyanont},
  {Vink{\'o}}, {Wang}, {Wheeler}, {Yang}, \& {Zheng}}]{Kwok2025}
{Kwok}, L.~A., {Liu}, C., {Jha}, S.~W., {et~al.} 2025, arXiv e-prints,
  arXiv:2510.09760, \dodoi{10.48550/arXiv.2510.09760}

\bibitem[{Leahy \& Ranasinghe(2016)}]{leahy_distance_2016}
Leahy, D.~A., \& Ranasinghe, S. 2016, The Astrophysical Journal, 817, 74,
  \dodoi{10.3847/0004-637X/817/1/74}

\bibitem[{Lee {et~al.}(2012)Lee, Ellison, \& Nagataki}]{lee_generalized_2012}
Lee, S.-H., Ellison, D.~C., \& Nagataki, S. 2012, The Astrophysical Journal,
  750, 156, \dodoi{10.1088/0004-637X/750/2/156}

\bibitem[{Lee {et~al.}(2014)Lee, Patnaude, Ellison, Nagataki, \&
  Slane}]{lee_reverse_2014}
Lee, S.-H., Patnaude, D.~J., Ellison, D.~C., Nagataki, S., \& Slane, P.~O.
  2014, The Astrophysical Journal, 791, 97, \dodoi{10.1088/0004-637X/791/2/97}

\bibitem[{Lee {et~al.}(2015)Lee, Patnaude, Raymond, Nagataki, Slane, \&
  Ellison}]{lee_modeling_2015}
Lee, S.-H., Patnaude, D.~J., Raymond, J.~C., {et~al.} 2015, The Astrophysical
  Journal, 806, 71, \dodoi{10.1088/0004-637X/806/1/71}

\bibitem[{Lee {et~al.}(2013)Lee, Slane, Ellison, Nagataki, \&
  Patnaude}]{lee_cr-hydro-nei_2013}
Lee, S.-H., Slane, P.~O., Ellison, D.~C., Nagataki, S., \& Patnaude, D.~J.
  2013, The Astrophysical Journal, 767, 20, \dodoi{10.1088/0004-637X/767/1/20}

\bibitem[{{Li} {et~al.}(2017){Li}, {Chu}, {Gruendl}, {Weisz}, {Pan}, {Points},
  {Ricker}, {Smith}, \& {Walter}}]{Li2017}
{Li}, C.-J., {Chu}, Y.-H., {Gruendl}, R.~A., {et~al.} 2017, \apj, 836, 85,
  \dodoi{10.3847/1538-4357/836/1/85}

\bibitem[{Liu {et~al.}(2023)Liu, Röpke, \& Han}]{liu_type_2023}
Liu, Z.-W., Röpke, F.~K., \& Han, Z. 2023, Research in Astronomy and
  Astrophysics, 23, 082001, \dodoi{10.1088/1674-4527/acd89e}

\bibitem[{Lopez {et~al.}(2009)Lopez, Ramirez-Ruiz, Badenes, Huppenkothen,
  Jeltema, \& Pooley}]{lopez_typing_2009}
Lopez, L.~A., Ramirez-Ruiz, E., Badenes, C., {et~al.} 2009, The Astrophysical
  Journal, 706, L106, \dodoi{10.1088/0004-637X/706/1/L106}

\bibitem[{Lopez {et~al.}(2011)Lopez, Ramirez-Ruiz, Huppenkothen, Badenes, \&
  Pooley}]{lopez_using_2011}
Lopez, L.~A., Ramirez-Ruiz, E., Huppenkothen, D., Badenes, C., \& Pooley, D.~A.
  2011, The Astrophysical Journal, 732, 114,
  \dodoi{10.1088/0004-637X/732/2/114}

\bibitem[{Maggi \& Acero(2017)}]{maggi_fe_2017}
Maggi, P., \& Acero, F. 2017, Astronomy and Astrophysics, 597, A65,
  \dodoi{10.1051/0004-6361/201629378}

\bibitem[{Maggi {et~al.}(2016)Maggi, Haberl, Kavanagh, Sasaki, Bozzetto,
  Filipović, Vasilopoulos, Pietsch, Points, Chu, Dickel, Ehle, Williams, \&
  Greiner}]{maggi_population_2016}
Maggi, P., Haberl, F., Kavanagh, P.~J., {et~al.} 2016, Astronomy and
  Astrophysics, 585, A162, \dodoi{10.1051/0004-6361/201526932}

\bibitem[{Mandal {et~al.}(2024)Mandal, Duffell, Polin, \&
  Milisavljevic}]{mandal_measurement_2024}
Mandal, S., Duffell, P.~C., Polin, A., \& Milisavljevic, D. 2024, The
  Astrophysical Journal, 972, 87, \dodoi{10.3847/1538-4357/ad5daa}

\bibitem[{Maoz {et~al.}(2014)Maoz, Mannucci, \&
  Nelemans}]{maoz_observational_2014}
Maoz, D., Mannucci, F., \& Nelemans, G. 2014, Annu. Rev. Astron. Astrophys.,
  52, 107, \dodoi{10.1146/annurev-astro-082812-141031}

\bibitem[{Margutti {et~al.}(2014)Margutti, Parrent, Kamble, Soderberg, Foley,
  Milisavljevic, Drout, \& Kirshner}]{margutti_no_2014}
Margutti, R., Parrent, J., Kamble, A., {et~al.} 2014, The Astrophysical
  Journal, 790, 52, \dodoi{10.1088/0004-637X/790/1/52}

\bibitem[{{Mart{\'\i}nez-Rodr{\'\i}guez}
  {et~al.}(2017){Mart{\'\i}nez-Rodr{\'\i}guez}, {Badenes}, {Yamaguchi},
  {Bravo}, {Timmes}, {Miles}, {Townsley}, {Piro}, {Mori}, {Andrews}, \&
  {Park}}]{Martinez2017}
{Mart{\'\i}nez-Rodr{\'\i}guez}, H., {Badenes}, C., {Yamaguchi}, H., {et~al.}
  2017, \apj, 843, 35, \dodoi{10.3847/1538-4357/aa72f8}

\bibitem[{Martínez-Rodríguez {et~al.}(2018)Martínez-Rodríguez, Badenes,
  Lee, Patnaude, Foster, Yamaguchi, Auchettl, Bravo, Slane, Piro, Park, \&
  Nagataki}]{martinez-rodriguez_chandrasekhar_2018}
Martínez-Rodríguez, H., Badenes, C., Lee, S.-H., {et~al.} 2018, ApJ, 865,
  151, \dodoi{10.3847/1538-4357/aadaec}

\bibitem[{McKinney(2010)}]{mckinney_data_2010}
McKinney, W. 2010, in Proceedings of th 9th {Python} in {Science} {Conference},
  Austin, Texas, 56--61, \dodoi{10.25080/Majora-92bf1922-00a}

\bibitem[{Meng \& Podsiadlowski(2017)}]{meng_common-envelope_2017}
Meng, X., \& Podsiadlowski, P. 2017, Monthly Notices of the Royal Astronomical
  Society, 469, 4763, \dodoi{10.1093/mnras/stx1137}

\bibitem[{Neumann {et~al.}(2024)Neumann, Tucker, Kochanek, Shappee, \&
  Stanek}]{neumann_echo_2024}
Neumann, K.~D., Tucker, M.~A., Kochanek, C.~S., Shappee, B.~J., \& Stanek,
  K.~Z. 2024, The Open Journal of Astrophysics, 7, 98,
  \dodoi{10.33232/001c.125246}

\bibitem[{Pannuti {et~al.}(2014)Pannuti, Kargaltsev, Napier, \&
  Brehm}]{pannuti_xmm-newton_2014}
Pannuti, T.~G., Kargaltsev, O., Napier, J.~P., \& Brehm, D. 2014, The
  Astrophysical Journal, 782, 102, \dodoi{10.1088/0004-637X/782/2/102}

\bibitem[{Patnaude \& Badenes(2017)}]{patnaude_supernova_2017}
Patnaude, D., \& Badenes, C. 2017, Supernova {Remnants} as {Clues} to {Their}
  {Progenitors}, \dodoi{10.1007/978-3-319-21846-5_98}

\bibitem[{Patnaude {et~al.}(2012)Patnaude, Badenes, Park, \&
  Laming}]{patnaude_origin_2012}
Patnaude, D.~J., Badenes, C., Park, S., \& Laming, J.~M. 2012, The
  Astrophysical Journal, 756, 6, \dodoi{10.1088/0004-637X/756/1/6}

\bibitem[{Patnaude {et~al.}(2009)Patnaude, Ellison, \&
  Slane}]{patnaude_role_2009}
Patnaude, D.~J., Ellison, D.~C., \& Slane, P. 2009, The Astrophysical Journal,
  696, 1956, \dodoi{10.1088/0004-637X/696/2/1956}

\bibitem[{Patnaude {et~al.}(2017)Patnaude, Lee, Slane, Badenes, Nagataki,
  Ellison, \& Milisavljevic}]{patnaude_impact_2017}
Patnaude, D.~J., Lee, S.-H., Slane, P.~O., {et~al.} 2017, The Astrophysical
  Journal, 849, 109, \dodoi{10.3847/1538-4357/aa9189}

\bibitem[{Patnaude {et~al.}(2010)Patnaude, Slane, Raymond, \&
  Ellison}]{patnaude_role_2010}
Patnaude, D.~J., Slane, P., Raymond, J.~C., \& Ellison, D.~C. 2010, The
  Astrophysical Journal, 725, 1476, \dodoi{10.1088/0004-637X/725/2/1476}

\bibitem[{{Prinja} {et~al.}(2000){Prinja}, {Ringwald}, {Wade}, \&
  {Knigge}}]{Prinja2000}
{Prinja}, R.~K., {Ringwald}, F.~A., {Wade}, R.~A., \& {Knigge}, C. 2000,
  \mnras, 312, 316, \dodoi{10.1046/j.1365-8711.2000.03111.x}

\bibitem[{Rakowski {et~al.}(2006)Rakowski, Badenes, Gaensler, Gelfand, Hughes,
  \& Slane}]{rakowski_can_2006}
Rakowski, C.~E., Badenes, C., Gaensler, B.~M., {et~al.} 2006, The Astrophysical
  Journal, 646, 982, \dodoi{10.1086/505018}

\bibitem[{Rest {et~al.}(2005)Rest, Suntzeff, Olsen, Prieto, Smith, Welch,
  Becker, Bergmann, Clocchiatti, Cook, Garg, Huber, Miknaitis, Minniti,
  Nikolaev, \& Stubbs}]{rest_light_2005}
Rest, A., Suntzeff, N.~B., Olsen, K., {et~al.} 2005, Nature, 438, 1132,
  \dodoi{10.1038/nature04365}

\bibitem[{Rest {et~al.}(2008)Rest, Matheson, Blondin, Bergmann, Welch,
  Suntzeff, Smith, Olsen, Prieto, Garg, Challis, Stubbs, Hicken, Modjaz,
  Wood-Vasey, Zenteno, Damke, Newman, Huber, Cook, Nikolaev, Becker, Miceli,
  Covarrubias, Morelli, Pignata, Clocchiatti, Minniti, \&
  Foley}]{rest_spectral_2008}
Rest, A., Matheson, T., Blondin, S., {et~al.} 2008, The Astrophysical Journal,
  680, 1137, \dodoi{10.1086/587158}

\bibitem[{Reynolds {et~al.}(2008)Reynolds, Borkowski, Green, Hwang, Harrus, \&
  Petre}]{reynolds_youngest_2008}
Reynolds, S.~P., Borkowski, K.~J., Green, D.~A., {et~al.} 2008, The
  Astrophysical Journal, 680, L41, \dodoi{10.1086/589570}

\bibitem[{Reynolds {et~al.}(2009)Reynolds, Borkowski, Green, Hwang, Harrus, \&
  Petre}]{reynolds_x-ray_2009}
---. 2009, The Astrophysical Journal, 695, L149,
  \dodoi{10.1088/0004-637X/695/2/L149}

\bibitem[{Reynolds {et~al.}(2007)Reynolds, Borkowski, Hwang, Hughes, Badenes,
  Laming, \& Blondin}]{reynolds_deep_2007}
Reynolds, S.~P., Borkowski, K.~J., Hwang, U., {et~al.} 2007, The Astrophysical
  Journal, 668, L135, \dodoi{10.1086/522830}

\bibitem[{Reynoso \& Goss(1999)}]{reynoso_new_1999}
Reynoso, E.~M., \& Goss, W.~M. 1999, The Astronomical Journal, 118, 926,
  \dodoi{10.1086/300990}

\bibitem[{Ruiter {et~al.}(2009)Ruiter, Belczynski, \&
  Fryer}]{ruiter_rates_2009}
Ruiter, A.~J., Belczynski, K., \& Fryer, C. 2009, The Astrophysical Journal,
  699, 2026, \dodoi{10.1088/0004-637X/699/2/2026}

\bibitem[{Ruiter \& Seitenzahl(2024)}]{ruiter_type_2024}
Ruiter, A.~J., \& Seitenzahl, I.~R. 2024, Type {Ia} supernova progenitors: a
  contemporary view of a long-standing puzzle,
  \dodoi{10.48550/arXiv.2412.01766}

\bibitem[{Sano {et~al.}(2022)Sano, Yamaguchi, Aruga, Fukui, Tachihara,
  Filipović, \& Rowell}]{sano_expanding_2022}
Sano, H., Yamaguchi, H., Aruga, M., {et~al.} 2022, The Astrophysical Journal,
  933, 157, \dodoi{10.3847/1538-4357/ac7465}

\bibitem[{Sarbadhicary {et~al.}(2019)Sarbadhicary, Chomiuk, Badenes, Tremou,
  Soderberg, \& Sjouwerman}]{sarbadhicary_two_2019}
Sarbadhicary, S.~K., Chomiuk, L., Badenes, C., {et~al.} 2019, The Astrophysical
  Journal, 872, 191, \dodoi{10.3847/1538-4357/ab027f}

\bibitem[{{Sarbadhicary} {et~al.}(2025){Sarbadhicary}, {Long}, {Raymond},
  {Sankrit}, {Egorov}, {Roman-Lopes}, {Blanc}, {Gelfand}, {Badenes}, {Drory},
  {Fern{\'a}ndez-Trincado}, {Garc{\'\i}a}, {Johnston}, {Jones}, {Katkov},
  {Kreckel}, {Li}, {Mej{\'\i}a-Narv{\'a}ez}, {M{\'e}ndez-Delgado},
  {Orozco-Duarte}, {Sanchez}, \& {Wong}}]{Sarbadhicary2025}
{Sarbadhicary}, S.~K., {Long}, K.~S., {Raymond}, J.~C., {et~al.} 2025, arXiv
  e-prints, arXiv:2507.08257, \dodoi{10.48550/arXiv.2507.08257}

\bibitem[{Sato {et~al.}(2019)Sato, Hughes, Williams, \&
  Morii}]{sato_genus_2019}
Sato, T., Hughes, J.~P., Williams, B.~J., \& Morii, M. 2019, The Astrophysical
  Journal, 879, 64, \dodoi{10.3847/1538-4357/ab24db}

\bibitem[{Scalzo {et~al.}(2014)Scalzo, Ruiter, \& Sim}]{scalzo_ejected_2014}
Scalzo, R.~A., Ruiter, A.~J., \& Sim, S.~A. 2014, Monthly Notices of the Royal
  Astronomical Society, 445, 2535, \dodoi{10.1093/mnras/stu1808}

\bibitem[{Schindelheim {et~al.}(2024)Schindelheim, Court, Badenes, Lee,
  Patnaude, García-Segura, \& Bravo}]{schindelheim_snr_2024}
Schindelheim, P., Court, T., Badenes, C., {et~al.} 2024, Research Notes of the
  American Astronomical Society, 8, 309, \dodoi{10.3847/2515-5172/ad9dd4}

\bibitem[{Siegel {et~al.}(2021)Siegel, Dwarkadas, Frank, \&
  Burrows}]{siegel_can_2021}
Siegel, J., Dwarkadas, V.~V., Frank, K.~A., \& Burrows, D.~N. 2021, The
  Astrophysical Journal, 922, 67, \dodoi{10.3847/1538-4357/ac2305}

\bibitem[{{Slane} {et~al.}(2014){Slane}, {Lee}, {Ellison}, {Patnaude},
  {Hughes}, {Eriksen}, {Castro}, \& {Nagataki}}]{Slane2014}
{Slane}, P., {Lee}, S.~H., {Ellison}, D.~C., {et~al.} 2014, \apj, 783, 33,
  \dodoi{10.1088/0004-637X/783/1/33}

\bibitem[{Stephenson \& Green(2002)}]{stephenson_historical_2002}
Stephenson, F.~R., \& Green, D.~A. 2002, International Series in Astronomy and
  Astrophysics, 5.
\newblock \url{https://ui.adsabs.harvard.edu/abs/2002ISAA....5.....S}

\bibitem[{Stritzinger {et~al.}(2006)Stritzinger, Mazzali, Sollerman, \&
  Benetti}]{stritzinger_consistent_2006}
Stritzinger, M., Mazzali, P.~A., Sollerman, J., \& Benetti, S. 2006, Astronomy
  and Astrophysics, 460, 793, \dodoi{10.1051/0004-6361:20065514}

\bibitem[{Takata {et~al.}(2016)Takata, Nobukawa, Uchida, Tsuru, Tanaka, \&
  Koyama}]{takata_x-ray_2016}
Takata, A., Nobukawa, M., Uchida, H., {et~al.} 2016, Publications of the
  Astronomical Society of Japan, 68, S3, \dodoi{10.1093/pasj/psv025}

\bibitem[{{Terwel} {et~al.}(2025){Terwel}, {Maguire}, {Brennan}, {Galbany},
  {Reusch}, {Schulze}, {Koivisto}, {Pursimo}, {Grund S{\o}rensen}, {D{\'\i}az
  Teodori}, {Guldberg Theil}, {Turkki}, {M{\"u}ller-Bravo}, {Burgaz}, {Kim},
  {Bloom}, {Graham}, {Kasliwal}, {Kulkarni}, {Masci}, {Purdum}, {Pyshna}, \&
  {Wold}}]{Terwel2025}
{Terwel}, J.~H., {Maguire}, K., {Brennan}, S.~J., {et~al.} 2025, \aap, 702,
  A21, \dodoi{10.1051/0004-6361/202555892}

\bibitem[{Theuns \& Jorissen(1993)}]{theuns_wind_1993}
Theuns, T., \& Jorissen, A. 1993, Monthly Notices of the Royal Astronomical
  Society, 265, 946, \dodoi{10.1093/mnras/265.4.946}

\bibitem[{{Toonen} {et~al.}(2018){Toonen}, {Perets}, \& {Hamers}}]{Toonen2018}
{Toonen}, S., {Perets}, H.~B., \& {Hamers}, A.~S. 2018, \aap, 610, A22,
  \dodoi{10.1051/0004-6361/201731874}

\bibitem[{Townsend(2009)}]{townsend_exact_2009}
Townsend, R. H.~D. 2009, The Astrophysical Journal Supplement Series, 181, 391,
  \dodoi{10.1088/0067-0049/181/2/391}

\bibitem[{Vink(2008)}]{vink_kinematics_2008}
Vink, J. 2008, The Astrophysical Journal, 689, 231, \dodoi{10.1086/592375}

\bibitem[{Vink {et~al.}(2006)Vink, Bleeker, van~der Heyden, Bykov, Bamba, \&
  Yamazaki}]{vink_x-ray_2006}
Vink, J., Bleeker, J., van~der Heyden, K., {et~al.} 2006, The Astrophysical
  Journal, 648, L33, \dodoi{10.1086/507628}

\bibitem[{{Vink} {et~al.}(2006){Vink}, {Bleeker}, {van der Heyden}, {Bykov},
  {Bamba}, \& {Yamazaki}}]{Vink2006}
{Vink}, J., {Bleeker}, J., {van der Heyden}, K., {et~al.} 2006, \apjl, 648,
  L33, \dodoi{10.1086/507628}

\bibitem[{{Vink} {et~al.}(2025){Vink}, {Agarwal}, {Bamba}, {Gu}, {Plucinsky},
  {Behar}, {Corrales}, {Foster}, {Fujimoto}, {Ichihashi}, {Ichikawa},
  {Katsuda}, {Matsumoto}, {Matsunaga}, {Mizuno}, {Mori}, {Murakami},
  {Nakajima}, {Sato}, {Sawada}, {Sonoda}, {Suzuki}, {Tateishi}, {Terada}, \&
  {Uchida}}]{Vink2025}
{Vink}, J., {Agarwal}, M., {Bamba}, A., {et~al.} 2025, arXiv e-prints,
  arXiv:2505.04691, \dodoi{10.48550/arXiv.2505.04691}

\bibitem[{Virtanen {et~al.}(2020)Virtanen, Gommers, Oliphant, Haberland, Reddy,
  Cournapeau, Burovski, Peterson, Weckesser, Bright, Van Der~Walt, Brett,
  Wilson, Millman, Mayorov, Nelson, Jones, Kern, Larson, Carey, Polat, Feng,
  Moore, VanderPlas, Laxalde, Perktold, Cimrman, Henriksen, Quintero, Harris,
  Archibald, Ribeiro, Pedregosa, Van~Mulbregt, {SciPy 1.0 Contributors},
  Vijaykumar, Bardelli, Rothberg, Hilboll, Kloeckner, Scopatz, Lee, Rokem,
  Woods, Fulton, Masson, Häggström, Fitzgerald, Nicholson, Hagen, Pasechnik,
  Olivetti, Martin, Wieser, Silva, Lenders, Wilhelm, Young, Price, Ingold,
  Allen, Lee, Audren, Probst, Dietrich, Silterra, Webber, Slavič, Nothman,
  Buchner, Kulick, Schönberger, De~Miranda~Cardoso, Reimer, Harrington,
  Rodríguez, Nunez-Iglesias, Kuczynski, Tritz, Thoma, Newville, Kümmerer,
  Bolingbroke, Tartre, Pak, Smith, Nowaczyk, Shebanov, Pavlyk, Brodtkorb, Lee,
  McGibbon, Feldbauer, Lewis, Tygier, Sievert, Vigna, Peterson, More, Pudlik,
  Oshima, Pingel, Robitaille, Spura, Jones, Cera, Leslie, Zito, Krauss,
  Upadhyay, Halchenko, \& Vázquez-Baeza}]{virtanen_scipy_2020}
Virtanen, P., Gommers, R., Oliphant, T.~E., {et~al.} 2020, Nat Methods, 17,
  261, \dodoi{10.1038/s41592-019-0686-2}

\bibitem[{Wang \& Han(2012)}]{wang_progenitors_2012}
Wang, B., \& Han, Z. 2012, New Astronomy Reviews, 56, 122,
  \dodoi{10.1016/j.newar.2012.04.001}

\bibitem[{Warren {et~al.}(2005)Warren, Hughes, Badenes, Ghavamian, McKee,
  Moffett, Plucinsky, Rakowski, Reynoso, \& Slane}]{warren_cosmic-ray_2005}
Warren, J.~S., Hughes, J.~P., Badenes, C., {et~al.} 2005, The Astrophysical
  Journal, 634, 376, \dodoi{10.1086/496941}

\bibitem[{Weaver {et~al.}(1977)Weaver, McCray, Castor, Shapiro, \&
  Moore}]{weaver_interstellar_1977}
Weaver, R., McCray, R., Castor, J., Shapiro, P., \& Moore, R. 1977, The
  Astrophysical Journal, 218, 377, \dodoi{10.1086/155692}

\bibitem[{{Webbink}(1984)}]{Webbink1984}
{Webbink}, R.~F. 1984, \apj, 277, 355, \dodoi{10.1086/161701}

\bibitem[{Williams {et~al.}(2013)Williams, Borkowski, Ghavamian, Hewitt, Mao,
  Petre, Reynolds, \& Blondin}]{williams_azimuthal_2013}
Williams, B.~J., Borkowski, K.~J., Ghavamian, P., {et~al.} 2013, The
  Astrophysical Journal, 770, 129, \dodoi{10.1088/0004-637X/770/2/129}

\bibitem[{Williams {et~al.}(2011)Williams, Blair, Blondin, Borkowski,
  Ghavamian, Long, Raymond, Reynolds, Rho, \& Winkler}]{williams_rcw_2011}
Williams, B.~J., Blair, W.~P., Blondin, J.~M., {et~al.} 2011, The Astrophysical
  Journal, 741, 96, \dodoi{10.1088/0004-637X/741/2/96}

\bibitem[{{Williams} {et~al.}(2014){Williams}, {Borkowski}, {Reynolds},
  {Ghavamian}, {Raymond}, {Long}, {Blair}, {Sankrit}, {Winkler}, \&
  {Hendrick}}]{Williams2014}
{Williams}, B.~J., {Borkowski}, K.~J., {Reynolds}, S.~P., {et~al.} 2014, \apj,
  790, 139, \dodoi{10.1088/0004-637X/790/2/139}

\bibitem[{Wood-Vasey \& Sokoloski(2006)}]{wood-vasey_novae_2006}
Wood-Vasey, W.~M., \& Sokoloski, J.~L. 2006, The Astrophysical Journal, 645,
  L53, \dodoi{10.1086/506179}

\bibitem[{Yamaguchi {et~al.}(2008)Yamaguchi, Koyama, Katsuda, Nakajima, Hughes,
  Bamba, Hiraga, Mori, Ozaki, \& Tsuru}]{yamaguchi_x-ray_2008}
Yamaguchi, H., Koyama, K., Katsuda, S., {et~al.} 2008, Publications of the
  Astronomical Society of Japan, 60, S141, \dodoi{10.1093/pasj/60.sp1.S141}

\bibitem[{Yamaguchi {et~al.}(2014{\natexlab{a}})Yamaguchi, Badenes, Petre,
  Nakano, Castro, Enoto, Hiraga, Hughes, Maeda, Nobukawa, Safi-Harb, Slane,
  Smith, \& Uchida}]{yamaguchi_discriminating_2014}
Yamaguchi, H., Badenes, C., Petre, R., {et~al.} 2014{\natexlab{a}}, ApJ, 785,
  L27, \dodoi{10.1088/2041-8205/785/2/L27}

\bibitem[{Yamaguchi {et~al.}(2014{\natexlab{b}})Yamaguchi, Eriksen, Badenes,
  Hughes, Brickhouse, Foster, Patnaude, Petre, Slane, \&
  Smith}]{yamaguchi_new_2014}
Yamaguchi, H., Eriksen, K.~A., Badenes, C., {et~al.} 2014{\natexlab{b}}, The
  Astrophysical Journal, 780, 136, \dodoi{10.1088/0004-637X/780/2/136}

\bibitem[{{Yamaguchi} {et~al.}(2015){Yamaguchi}, {Badenes}, {Foster}, {Bravo},
  {Williams}, {Maeda}, {Nobukawa}, {Eriksen}, {Brickhouse}, {Petre}, \&
  {Koyama}}]{Yamaguchi2015}
{Yamaguchi}, H., {Badenes}, C., {Foster}, A.~R., {et~al.} 2015, \apjl, 801,
  L31, \dodoi{10.1088/2041-8205/801/2/L31}

\bibitem[{Zhang {et~al.}(2023)Zhang, Zhou, Chen, Zhang, Zhong, Zhou, Zhang, \&
  Vink}]{zhang_molecular_2023}
Zhang, Q.-Q., Zhou, P., Chen, Y., {et~al.} 2023, The Astrophysical Journal,
  952, 107, \dodoi{10.3847/1538-4357/acd636}

\end{thebibliography}
\bibliographystyle{aasjournal}



\end{document}